\documentclass[reprint,twocolumn]{revtex4}%
\usepackage{amsfonts}
\usepackage{amsmath}
\usepackage{amssymb}
\usepackage{charter}
\usepackage{graphicx}%
\providecommand{\U}[1]{\protect \rule{.1in}{.1in}}
\begin{document}
\title{Improvement of phase sensitivity in SU(1,1) interferometer via a Kerr nonlinear}
\author{{\small Shoukang Chang}$^{{\small 1}}${\small , Wei Ye}$^{{\small 1,4}}%
${\small , Huan Zhang}$^{{\small 1}}${\small , Liyun Hu}%
$^{1,*}{\small \thanks{Corresponding author. Email: hlyun2008@126.com}}${\small ,
Jiehui Huang}$^2{\small \thanks{Corresponding author. Email:
jiehuihuang@126.com}},${\small and} {\small Sanqiu Liu}$%
^{1,3}{\small \thanks{Corresponding author. Email: sqliu@ncu.edu.cn}}$}
\affiliation{$^{1}${\small Center for Quantum Science and Technology, Jiangxi Normal
University, Nanchang 330022, China}}
\affiliation{$^{2}${\small School of Mathematics, Physics and Statistics, Shanghai
University of Engineering Science, Shanghai 201620, China}}
\affiliation{$^{3}${\small Department of physics, Nanchang University, 330031, China}}
\affiliation{$^{4}${\small School of Computer Science and Engineering, Central South
University, Changsha 410083, China}}

\begin{abstract}
We propose a theoretical scheme to enhance the phase sensitivity by
introducing a Kerr nonlinear phase shift into the traditional SU(1,1)
interferometer with a coherent state input and homodyne detection. We
investigate the realistic effects of photon losses on phase sensitivity and
quantum Fisher information. The results show that compared with the linear
phase shift in SU(1,1) interferometer, the Kerr nonlinear case can not only
enhance the phase sensitivity and quantum Fisher information, but also
significantly suppress the photon losses. We also observe that at the same
accessible parameters, internal losses have a greater influence on the phase
sensitivity than the external ones. It is interesting that, our scheme shows
an obvious advantage of low-cost input resources to obtain higher phase
sensitivity and larger quantum Fisher information due to the introduction of
nonlinear phase element.


\end{abstract}
\maketitle

\section{Introduction}

Quantum metrology has a close relation to various-important information areas,
such as Bose-Einstein condensate \cite{1,2,3}, gravitational wave detection
\cite{4,5}, and quantum imaging \cite{6,7,8}. It has been widely concerned and
highly developed in recent years. To meet the high precision demand, all kinds
of optical interferometers have been proposed. For instance, as a general
model, the Mach--Zehnder interferometer (MZI) has been used to be an essential
tool to provide insight into tiny variations on phase shift \cite{9,10,11}.

In order to improve the precise measurement, generally speaking, we can focus
on the following three stages \cite{12}: probe generation \cite{13,14,15,16},
probe modification \cite{17,18,19,20,21} and probe readout \cite{22,23}, as
illustrated in Fig. 1(a). In particular, for the probe generation, the phase
sensitivity is always confined to the standard quantum limit (SQL) when the
classical resources are injected into the input ports of the MZI. To surpass
the SQL, non-classical quantum states have widely used as the input of the
MZI, such as entangled states \cite{13}, twin Fock states \cite{24}, and NOON
states \cite{25}, by which the Heisenberg limit (HL) even can be reached
\cite{26,27}. Although the usage of nonclassical states can greatly improve
the phase sensitivity of optical interferometers, these states with large
average photon numbers are not only more difficult to prepare, but also very
fragile especially in the presence of environmental interferences
\cite{28,29,30}. Thus, from the viewpoint of resource theory, it will be a
challenging task with simple input state (such as coherent state (CS)) to
further improve the precision of measurement, especially in the realistic case.

On the other hand, many efforts have been paid to the stage of probe
modification, especially when Yurke \emph{et al}. first proposed an SU(1,1)
interferometer with a linear phase shift \cite{20}. In this system, the active
nonlinear optical devices, such as four-wave mixers (FWMs) and optical
parametric amplifiers (OPAs), are used instead of the passive linear beam
splitters (BSs) used in the conventional MZI \cite{31,32,33,34,35,36,37}. To
beat the SQL, Plick \emph{et al}. applied strong CS as the inputs into the
SU(1,1) interferometer \cite{33}. Subsequently, Li \emph{et al}. proposed a
scheme of reaching HL sensitivity via a squeezed vacuum state (SVS) plus a CS
with the homodyne detection \cite{31}. It is interesting that, Hudelist
\emph{et al}. pointed out that the signal-to-noise ratio (SNR) of the SU(1,1)
interferometer is about 4.1 dB higher than that of the MZI under the same
phase-sensing intensity \cite{38}. This point may be one of reasons focusing
on this interferometer. Actually, this implies the role of nonlinear process
for improving the precision.

Expect the SU(1,1) nonlinear process, the nonlinear phase shifts have also
been proposed for enhancing the phase estimation, which can be viewed as
another way of the probe modification \cite{9,39,40,41,42}. For instance, in
the traditional MZI, Zhang \textit{et al.} investigated the phase estimation
by replacing linear phase shift by nonlinear one \cite{9} using a CS and
parity measurement. Jiao \textit{et al.} proposed an improved protocol of
nonlinear phase estimation by inserting a nonlinear phase shift into the
traditional MZI, with active correlation output readout and homodyne detection
\cite{39}. More recently, Chang \textit{et al}. suggested a scheme for
enhancing phase sensitivity by introducing nonlinear phase shifter to the
modified interferometer consisting of a balanced BS and an optical parameter
amplifier (OPA) \cite{41}. It is shown that the OPA potential can be
stimulated by nonlinear phase shifter, which is absent for the case of linear
phase shifter. In addition, the estimation of nonlinear phase has also lots of
applications, such as the third-order susceptibility of the Kerr medium
\cite{39}, phase-sensitive amplifiers \cite{43,44} and nonclassical quantum
state preparations \cite{45,46}. These results show that the nonlinear optical
devices can be considered as powerful tools to effectively achieve both high
accuracy and sensitivity. However, on one hand, the researches on nonlinear
phase estimation are not as systematic as those on linear one. On the other,
most works on the phase precision are based on either specific measurement,
especially in the presence of photon loss, or direct calculation of quantum
Fisher information for ideal case.

In this paper, we mainly focus on the nonlinear phase estimation of a Kerr
SU(1,1) (KSU(1,1)) interferometer by replacing the linear phase shift with the
Kerr nonlinear one, together with CS plus vacuum state (VS) (denoted as
$\left \vert \alpha \right \rangle _{a}\otimes \left \vert 0\right \rangle _{b}$) as
inputs and homodyne detection. Both the phase sensitivity and the quantum
Fisher information (QFI) are analytically investigated with and without photon
losses. The phase sensitivity can beat the SQL and approach the HL. Compared
to the traditional SU(1,1) interferometer with a linear phase and other
inputs resources including CS+CS and SVS+CS, our scheme presents much QFI and
higher phase sensitivity closer to the quantum Cram\'{e}r-Rao (QCRB)
\cite{34,47,48,49,50}. From the viewpoint of resource theory, CS+VS can be
seen as the most simple and easily available input, thus our scheme has an
obvious advantage of low-cost input by inserting nonlinear phase shift into
the SU(1,1) interferometer.

The remainder of this paper is arranged as follows. Section II introduces the
KSU(1,1) interferometer in our scheme. In Section III, we  investigate the
phase sensitivity of the output signal with homodyne detection, and compare
them with the conventional SU(1,1) interferometer. In Section IV, we get the
QFI of Kerr nonlinear phase shift for the KSU(1,1) interferometer by invoking
the characteristic function (CF) approach. In Section V, we discuss the
effects of photon losses on both phase sensitivity and QFI of Kerr nonlinear
phase shift, respectively. Conclusions are made in the last section.
\begin{figure}[ptb]
\centering \includegraphics[width=1.0\columnwidth]{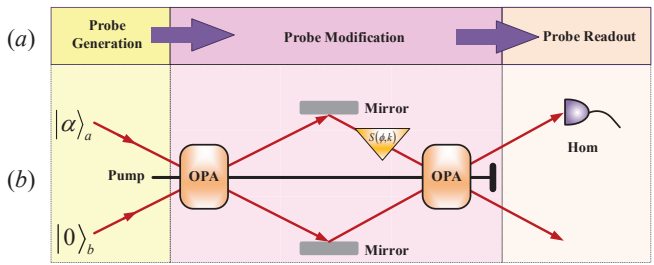}\caption{{}(Color
online) (a) The general process of estimating an unknown parameter $\phi$. (b)
Schematic diagram of SU(1,1) interferometer with a nonlinear phase shifter
$S\left(  \phi,k\right)  .$ The two input ports of this interferometer are a
coherent state $\left \vert \alpha \right \rangle _{a}$ and a vacuum state
$\left \vert 0\right \rangle _{b},$ respectively. OPA is an optical parametric
amplifier and Hom is a homodyne detection. }%
\end{figure}

\section{Nonlinear phase estimation model}

Let us begin with the description of nonlinear phase estimation model, as
shown in Fig. 1(b). The nonlinear interferometer consists of two OPAs (or
FWMs) and a Kerr-type medium. Here we consider a CS $\left \vert \alpha
\right \rangle _{a}$ with $\alpha=\left \vert \alpha \right \vert e^{i\theta
_{\alpha}}$\ and a VS $\left \vert 0\right \rangle _{b}$ as the inputs in mode
$a$ and mode $b$, respectively. Thus the probing state can be shown as
$\left \vert \psi_{in}\right \rangle =$ $\left \vert \alpha \right \rangle
_{a}\otimes \left \vert 0\right \rangle _{b}.$ After going through the first OPA,
the resulting state is given by $\hat{S}\left(  \xi_{1}\right)  \left \vert
\psi_{in}\right \rangle $ i.e., a two-mode squeezed CS, where the operator
$\hat{S}\left(  \xi_{1}\right)  =\exp(\xi_{1}^{\ast}\hat{a}\hat{b}-\xi_{1}%
\hat{a}^{\dagger}\hat{b}^{\dagger})$ represents the OPA process, $\xi
_{1}=g_{1}e^{i\theta_{1}}$, with a gain factor $g_{1}$ and a phase shift
$\theta_{1}$ and $\hat{a}(\hat{a}^{\dagger}),\hat{b}(\hat{b}^{\dagger})$ are
the annihilation (creation) operators, for modes $a$ and $b$, respectively.
For simplicity, we assume that the Kerr-type medium is inset into the path $b$
between the first and second OPAs to generate a nonlinear phase shift $\phi$
to be estimated.

After the interaction between the state $\hat{S}\left(  \xi_{1}\right)
\left \vert \psi_{in}\right \rangle $ with the Kerr-type medium, the
corresponding modified state becomes
\begin{equation}
\left \vert \psi_{\phi}\right \rangle =\hat{S}\left(  \phi,k\right)  \hat
{S}\left(  \xi_{1}\right)  \left \vert \psi_{in}\right \rangle , \label{0}%
\end{equation}
depending on the phase parameter $\phi$, where
\begin{equation}
\hat{S}\left(  \phi,k\right)  =e^{i\phi(\hat{b}^{\dagger}\hat{b})^{k}},
\label{1}%
\end{equation}
is the nonlinear phase shift operator and the exponent $k$ is the order of the
nonlinearity. In particular, for the case of $k=1$, $\hat{S}\left(
\phi,1\right)  =e^{i\phi \hat{b}^{\dagger}\hat{b}}$ just reduces to the linear
phase shift, while for the case of $k=2$, $\hat{S}\left(  \phi,2\right)
=e^{i\phi(\hat{b}^{\dagger}\hat{b})^{2}}$ corresponds to Kerr nonlinear case.
Throughout this paper, we only consider both Kerr-type nonlinear medium and
linear one by taking $k=1,2$, respectively.

After the second OPA, the final output state is given by%
\begin{equation}
\left \vert \psi_{out}\right \rangle =\hat{S}\left(  \xi_{2}\right)  \hat
{S}\left(  \phi,k\right)  \hat{S}\left(  \xi_{1}\right)  \left \vert \psi
_{in}\right \rangle ,\label{2}%
\end{equation}
where $\hat{S}\left(  \xi_{2}\right)  =\exp(\xi_{2}^{\ast}\hat{a}\hat{b}%
-\xi_{2}\hat{a}^{\dagger}\hat{b}^{\dagger}),$ with $\xi_{2}=g_{2}%
e^{i\theta_{2}},$ is a two-mode squeezing operator, corresponding to the
second OPA process. The Kerr nonlinear phase shifter $\hat{S}\left(
\phi,2\right)  $ satisfying the transformation relation:
\begin{equation}
\hat{S}^{\dagger}\left(  \phi,2\right)  \hat{b}^{\dagger}\hat{S}\left(
\phi,2\right)  =e^{-i\phi}\hat{b}^{\dagger}e^{-i2\phi \hat{b}^{\dagger}\hat{b}%
},\label{2b}%
\end{equation}
which is useful for the calculation of phase sensitivity. One can refer to
Appendix A about more details of derivation for this relation. Finally, the
homdyne detection is performed on mode $a$, so that one can read information
about the value of $\phi$. \begin{figure}[ptb]
\label{Fig2}
\centering \includegraphics[width=0.72\columnwidth]{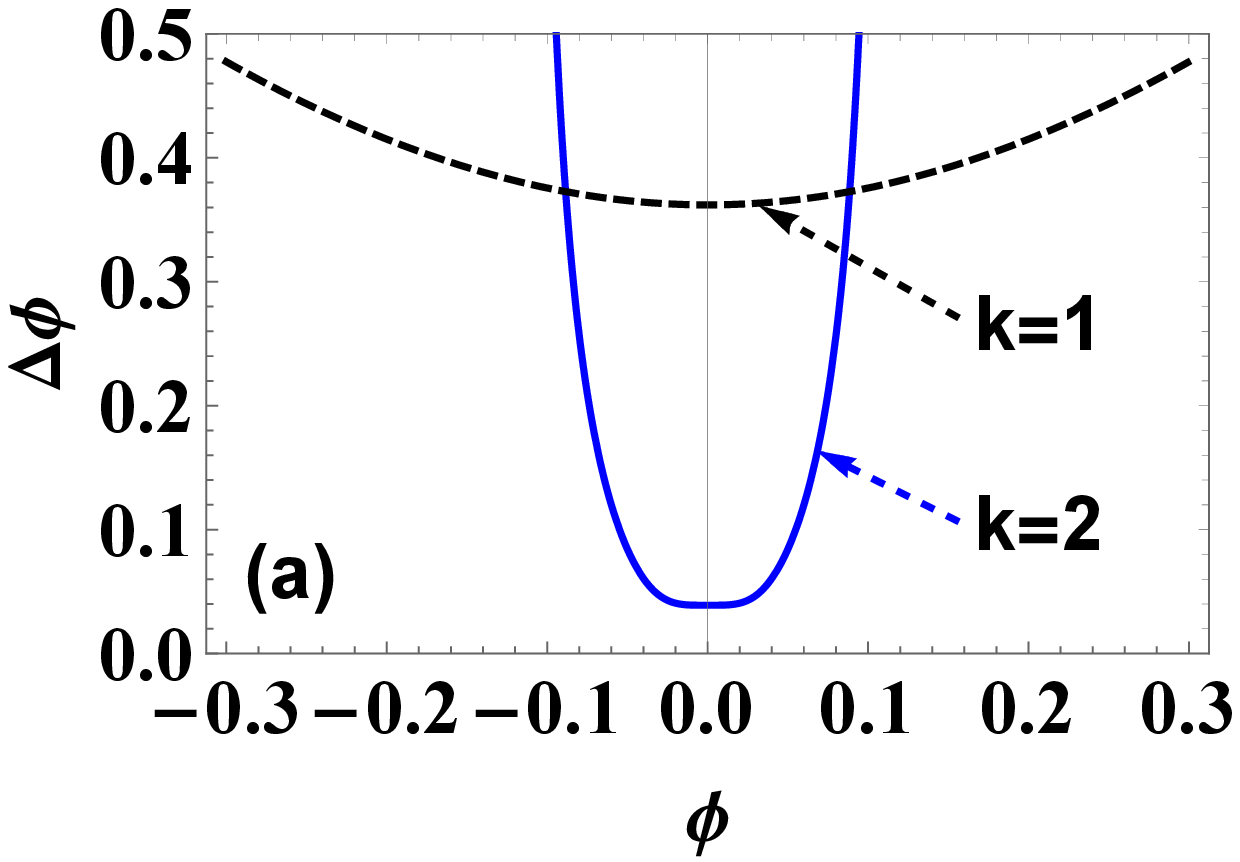}\newline%
\includegraphics[width=0.72\columnwidth]{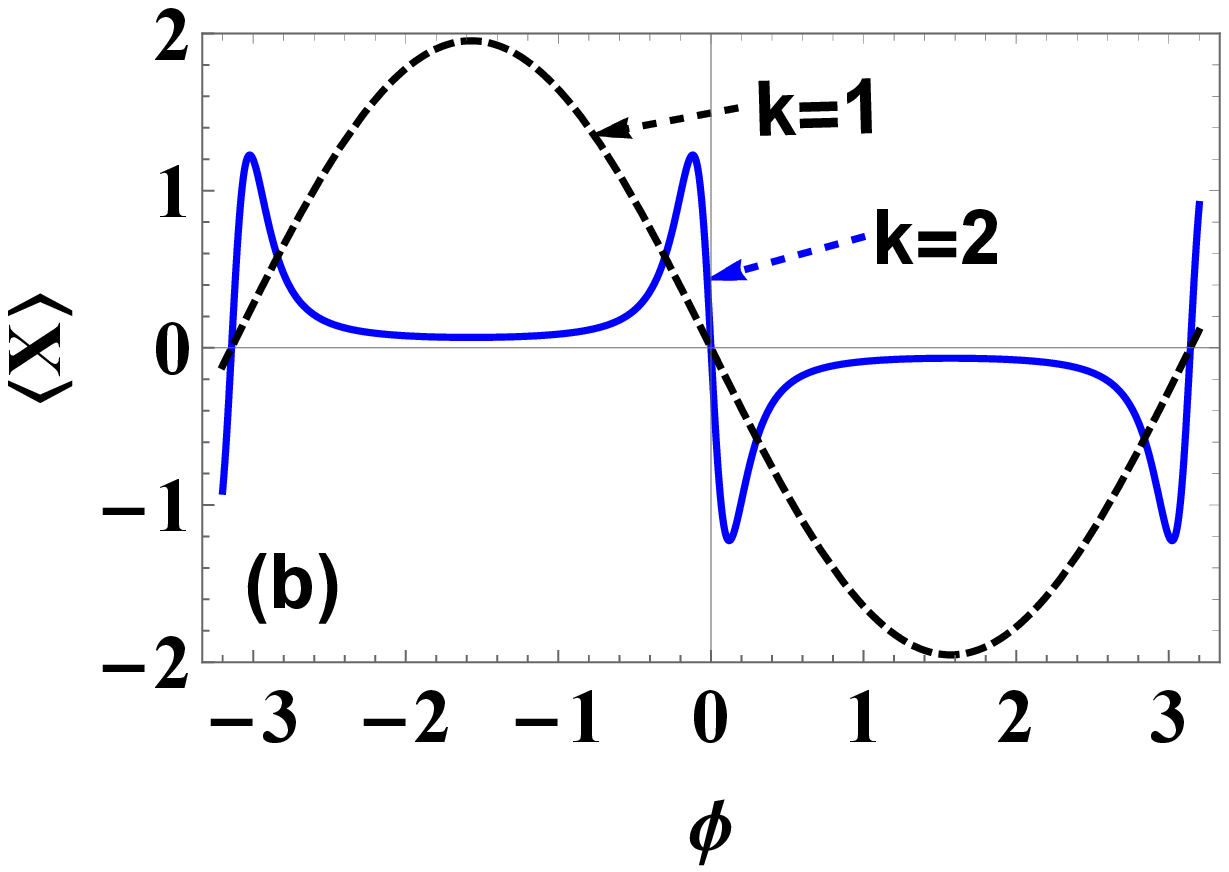} \newline \caption{{}(Color
online) (a) Phase sensitivity based on homodyne detection and (b) the output
signal as a function of $\phi$ with $g=1$, $\left \vert \alpha \right \vert =1$,
and $\theta_{\alpha}=\frac{\pi}{2}$. The black-dashed line and the blue-solid
line correspond to the linear phase shift ($k=1$) and the Kerr nonlinear phase
shift ($k=2$), respectively.}%
\end{figure}\begin{figure}[ptb]
\label{Fig3}
\centering \includegraphics[width=0.72\columnwidth]{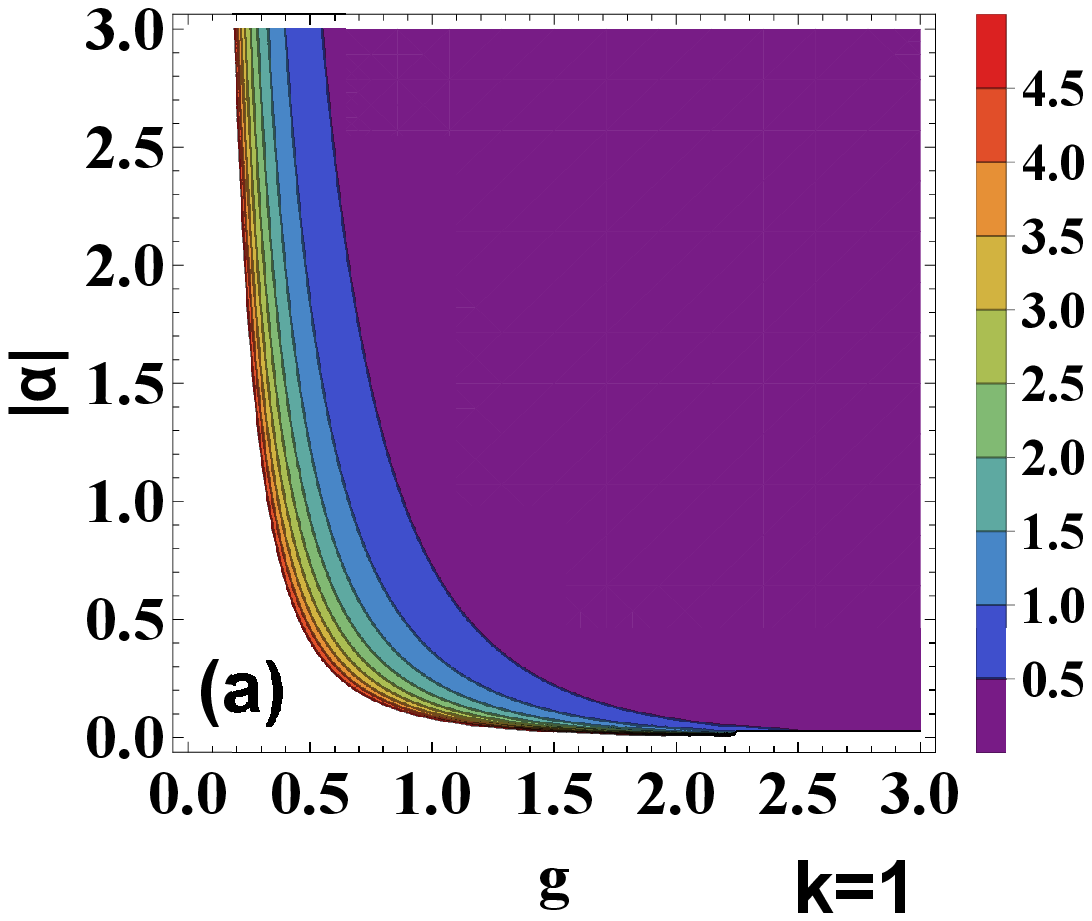}\newline%
\includegraphics[width=0.73\columnwidth]{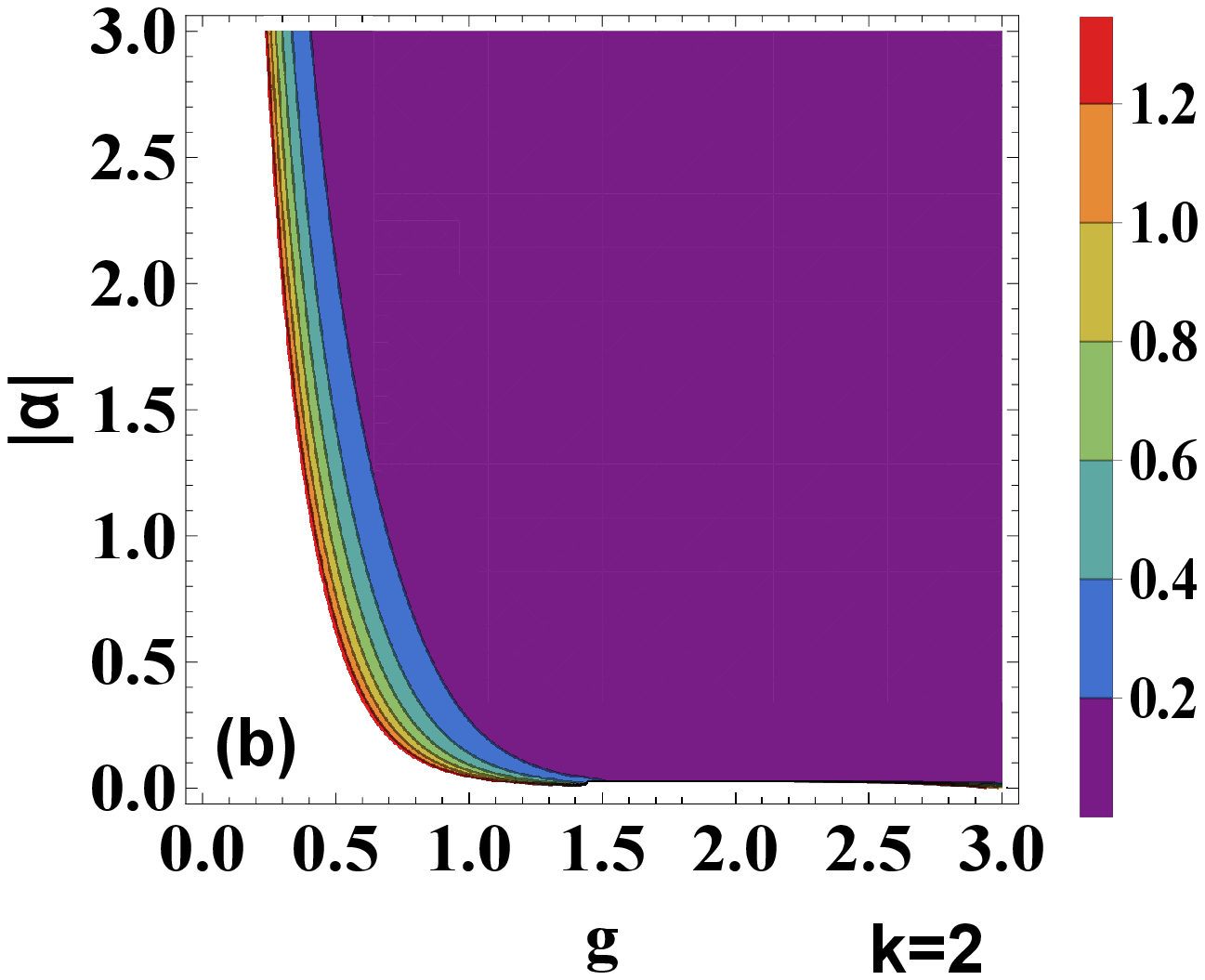} \newline \caption{{}(Color
online) Phase sensitivity based on homodyne detection as a function of the
gain factor $g$ and the coherent amplitude $\left \vert \alpha \right \vert $
with $\theta_{\alpha}=\pi/2$ and $\phi=0$ for (a) the linear phase shift
($k=1$), and (b) the Kerr nonlinear phase shift ($k=2$), respectively.}%
\end{figure}

\section{Phase sensitivity via homodyne detection}

Next, we investigate the phase sensitivity of nonlinear phase. For this
purpose, we need to choose a special detection method for the readout of phase
information at the final output port. Actually, there are many different
detection methods, such as homodyne detection \cite{11,31,32}, intensity
detection \cite{33,34,47}, and parity detection \cite{27,35,36}. Each way of
measurement has its own advantages. For example, the parity detection has been
proved to be an optimal detection for linear phase estimation in lots of
schemes \cite{27,51}. Compared with both intensity and parity detections,
however, homodyne detection can be easily realized with current technologies,
thereby playing a key role in the field of continuous-variable quantum key
distribution (CV-QKD) \cite{52,53,54,55,56}. Therefore, in our scheme the
homodyne detection is employed on mode $a$ at one of output port to estimate
the phase parameter $\phi,$ where the detected variable is the amplitude
quadrature $\hat{X}$, i.e.,
\begin{equation}
\hat{X}=(\hat{a}+\hat{a}^{\dagger})/\sqrt{2}. \label{3}%
\end{equation}

Based on the error propagation formula, the phase sensitivity can be
calculated by%
\begin{equation}
\Delta \phi=\frac{\sqrt{\Delta^{2}\hat{X}}}{\left \vert \partial \left \langle
\hat{X}\right \rangle /\partial \phi \right \vert }, \label{4}%
\end{equation}
with $\Delta^{2}\hat{X}=\left \langle \hat{X}^{2}\right \rangle -\left \langle
\hat{X}\right \rangle ^{2}.$ According to Eq. (\ref{4}), for an arbitrary value
of $\phi$, the corresponding phase sensitivity can be analytically derived.
For simplicity, the corresponding expression is not shown here. One can refer
to Appendix B for more details. In the following discussions, we assume that
the KSU(1,1) interferometer is in a balanced situation, i.e., $\theta
_{2}-\theta_{1}=\pi$ and $g_{1}=g_{2}=g.$

In Fig. 2(a), we show the phase sensitivity changing with $\phi$ for the
linear ($k=1$) and nonlinear ($k=2$) phase shifts when given parameters
$g=1,|\alpha|=1,$and $\theta_{\alpha}=\pi/2.$ It is clearly seen that, for
both cases above, the minimum value of the phase sensitivity can be achieved
at the optimal point $\phi=0.$ In addition, the phase sensitivity for $k=2$ is
always significantly superior to that for $k=1$ around the optimal point. This
implies that the nonlinear phase shift can be further used to improve the
phase sensitivity, compared to the linear one. The reason lies in that the
introduction of nonlinear phase shift increases the slope $\partial
\left \langle \hat{X}\right \rangle /\partial \phi$ of the output signal
$\left \langle \hat{X}\right \rangle $ at $\phi=0,$ which leads to the increase
of the denominator in Eq. (\ref{4}), as shown in Fig. 2(b). In fact, this
point will be clear by deriving the phase sensitivity $\Delta \phi_{k}$ for
$k=1,2$ at $\phi=0$, which are given by
\begin{align}
\Delta \phi_{1}  &  =\frac{1}{\sqrt{N_{\alpha}}N_{OPA}},\label{5a}\\
\Delta \phi_{2}  &  =\frac{\Delta \phi_{1}}{1+N_{OPA}\left(  N_{\alpha
}+2\right)  }, \label{5}%
\end{align}
where $\Delta \phi_{1}$ represents the phase sensitivity of the traditional
SU(1,1) interferometer with a linear phase shift $k=1,$ $N_{\alpha}=\left \vert
\alpha \right \vert ^{2}$ is the mean photon number of the input coherent state,
and $N_{OPA}=2\sinh^{2}g$ is the total photon number after the first OPA with
vacuum inputs. Obviously, $\Delta \phi_{2}>\Delta \phi_{1}.$

Moreover, from Fig. 2(b) it is shown that the peak width for the case of $k=2$
is narrower than that for the case of $k=1.$ In this sense, the nonlinear
phase shift can also improve supperresolution compared to the linear one. On
the other hand, to display the effects of both the coherence amplitude
$\left \vert \alpha \right \vert $ and the gain factor $g$ on the phase
sensitivity, at the optimal point $\phi=0,$ we also plot the phase sensitivity
as a function of $\left \vert \alpha \right \vert $ and $g$ with different values
of $k=1,2$, in Figs. 3(a) and 3(b), respectively. It is found that, the values
of $\Delta \phi$ rapidly decrease with the increase of $\left \vert
\alpha \right \vert $ and $g$, especially for the case of $k=2.$
\begin{figure}[ptb]
\label{Fig4}
\centering \includegraphics[width=0.72\columnwidth]{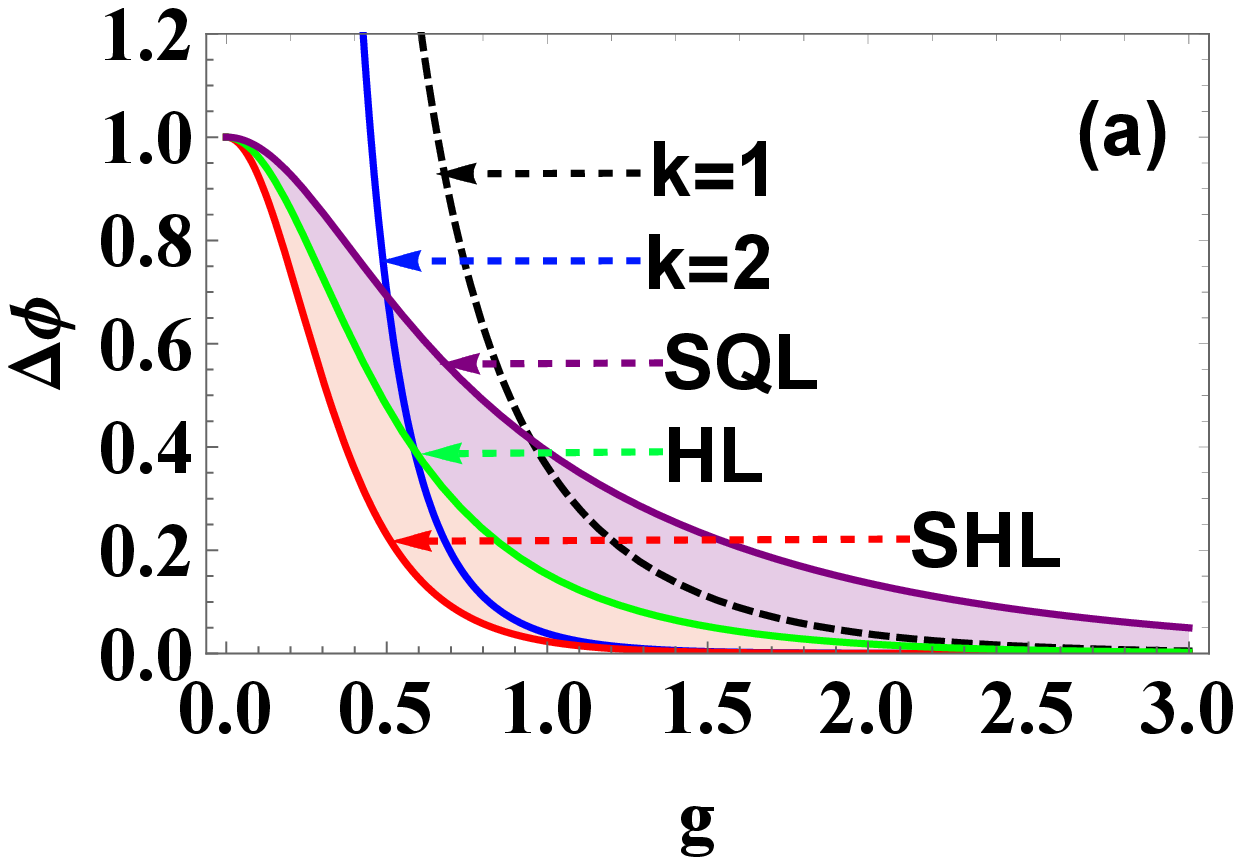}\newline%
\includegraphics[width=0.73\columnwidth]{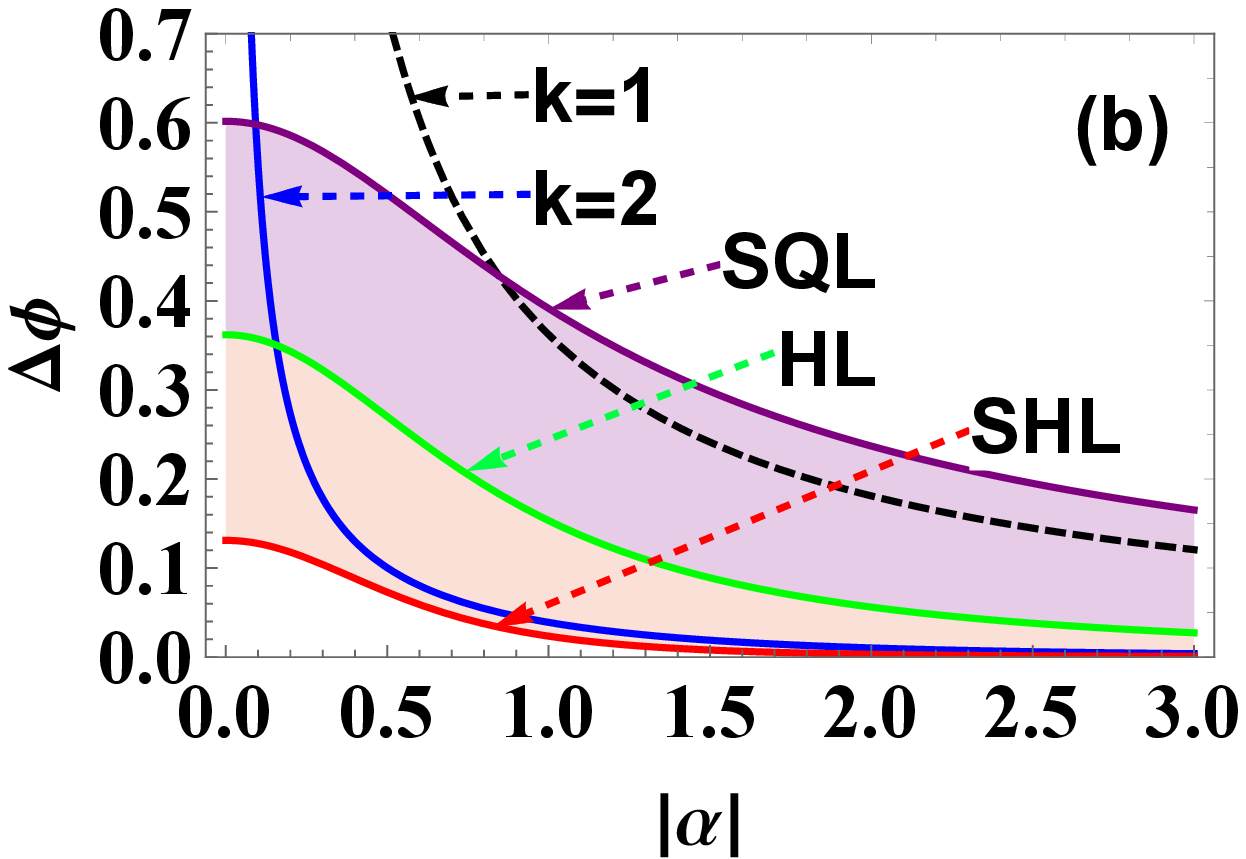} \newline \caption{(Color
online) Phase sensitivity based on homodyne detection as a function of (a) the
gain factor $g$ with $\left \vert \alpha \right \vert =1$. (b) the coherent
amplitude $\left \vert \alpha \right \vert $ with $g=1$ ($\theta_{\alpha}=\pi/2$
)$.$ The black-dashed and blue-solid lines are respectively the linear phase
shift ($k=1$), and the Kerr nonlinear phase shift ($k=2$), while the purple-,
green- and red-solid lines respectively correspond to the SQL, HL and SHL.}%
\end{figure}\begin{figure}[ptb]
\label{Fig5}
\centering \includegraphics[width=0.7\columnwidth]{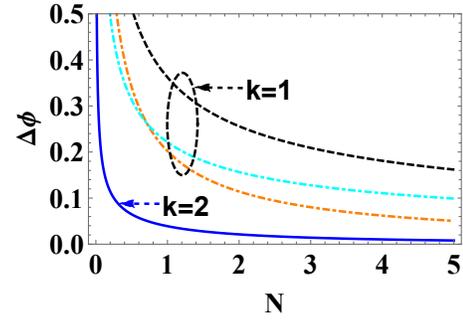}\caption{{}(Color
online) Phase sensitivity based on homodyne detection as a function of the
total average photon number N of input state with $g=1.$ The orange- and
cyan-dot-dashed lines respectively correspond to a squeezed vacuum state
combined with a coherent states input and two coherent states with the linear
phase shift ($k=1$), as shown in Ref. \cite{31,32}; while the black-dashed and
blue-solid lines are respectively the linear phase shift ($k=1$), and the Kerr
nonlinear phase shift ($k=2$). }%
\end{figure}\

To further show the advantage of our scheme, we also make a comparison about
phase sensitivities, involving the SQL ($\Delta \phi \thicksim1/\sqrt{N_{Total}%
}$), the HL ($\Delta \phi \thicksim1/N_{Total}$) and the Super HL (SHL)
($\Delta \phi \thicksim1/N_{Total}^{2}$), as shown in Fig. 4.\ Note that
$N_{Total}=\left \langle \psi_{in}\right \vert \hat{S}^{\dagger}\left(  \xi
_{1}\right)  (\hat{a}^{\dagger}\hat{a}+\hat{b}^{\dagger}\hat{b})\hat{S}\left(
\xi_{1}\right)  \left \vert \psi_{in}\right \rangle $ is the total mean photon
number inside the KSU(1,1) interferometer. From Fig. 4, the black dashed line
is the sensitivity performance of the linear phase shift, which can only break
through SQL and is always surpassed by that of the Kerr nonlinear phase shift.
In particular, the latter can break both the SQL and the HL, but cannot beat
the SHL. The reason may be that adopting the Kerr nonlinear phase shift to
effectively improve the maximum amount of information about the unknown phase
shift $\phi$ can result in a transition from HL ($\Delta \phi \thicksim
1/N_{Total}$) to SHL ($\Delta \phi \thicksim1/N_{Total}^{2}$), as described in
Refs. \cite{9,57,58,59,60,61,62}. Nevertheless, the common feature of these
two is that the phase sensitivity increases with the increase of $\left \vert
\alpha \right \vert $ and $g$.

Next, we make a comparison about the performance of phase sensitivity. In Fig.
5, the phase sensitivity is plotted as a function of the total average input photon
number $N$ with several different input resources, including $\left \vert
\alpha \right \rangle _{a}\otimes \left \vert \beta \right \rangle _{b}$,
$\left \vert \alpha \right \rangle _{a}\otimes \left \vert \varsigma,0\right \rangle
_{b}$ and $\left \vert \alpha \right \rangle _{a}\otimes \left \vert 0\right \rangle
_{b}$ used in the traditional SU(1,1) interferometer ($k=1$) and $\left \vert
\alpha \right \rangle _{a}\otimes \left \vert 0\right \rangle _{b}$ used in the
KSU(1,1) interferometer ($k=2$). It is shown that, for the traditional case,
the phase sensitivity with the input $\left \vert \alpha \right \rangle
_{a}\otimes \left \vert 0\right \rangle _{b}$ performs the worst among these
three inputs \cite{31,32}. This implies that the introduction of coherence
amplitude $\left \vert \alpha \right \vert $ or squeezing parameter $\varsigma$
is beneficial for the phase sensitivity improvement. However, for the KSU(1,1)
interferometer, it is interesting that the best phase sensitivity can be
achieved only using the simplest input $\left \vert \alpha \right \rangle
_{a}\otimes \left \vert 0\right \rangle _{b}$, which is significantly superior to
those inputs in the traditional SU(1,1) interferometer. That is to say, for a
simple input with less energy and less resources, the phase sensitivity can be
further improved by introducing Kerr nonlinear phase. \begin{figure}[ptb]
\label{Fig6} \centering \includegraphics[width=0.72\columnwidth]{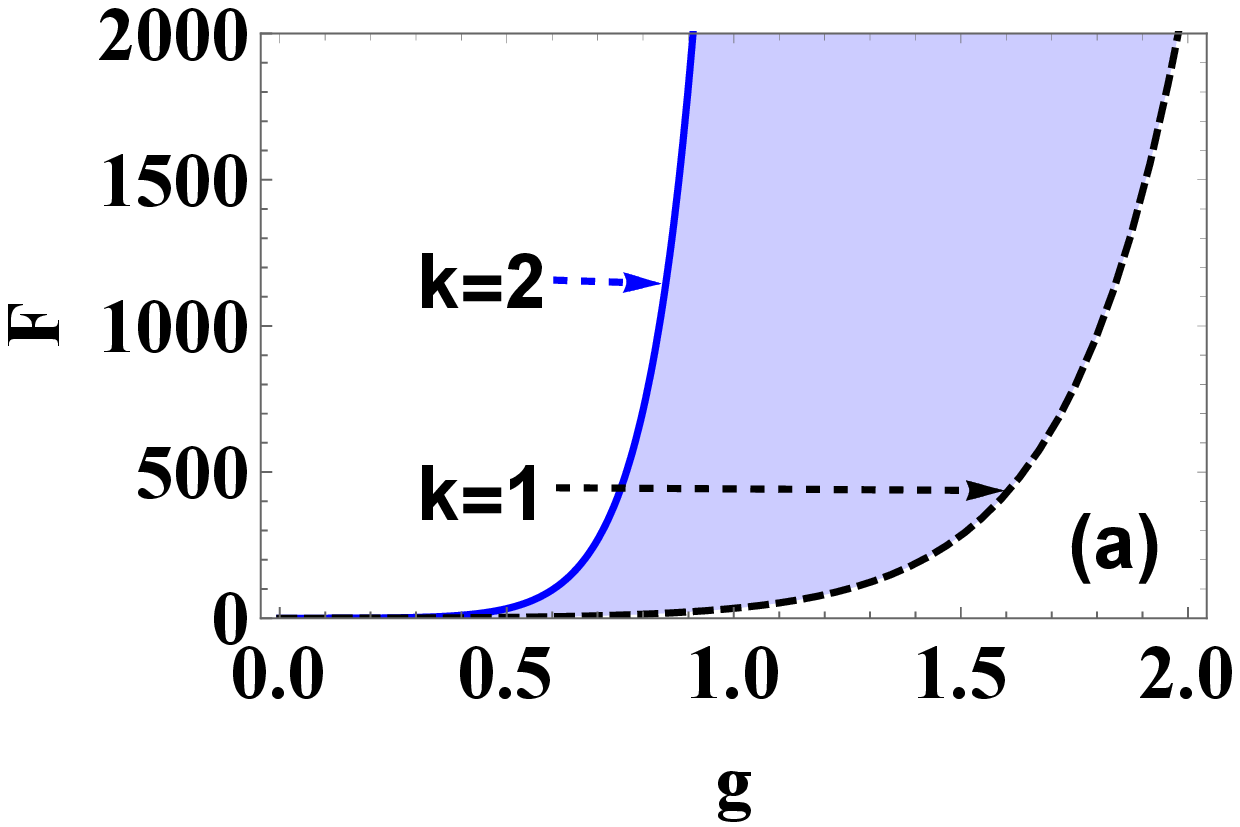}
\newline \includegraphics[width=0.73\columnwidth]{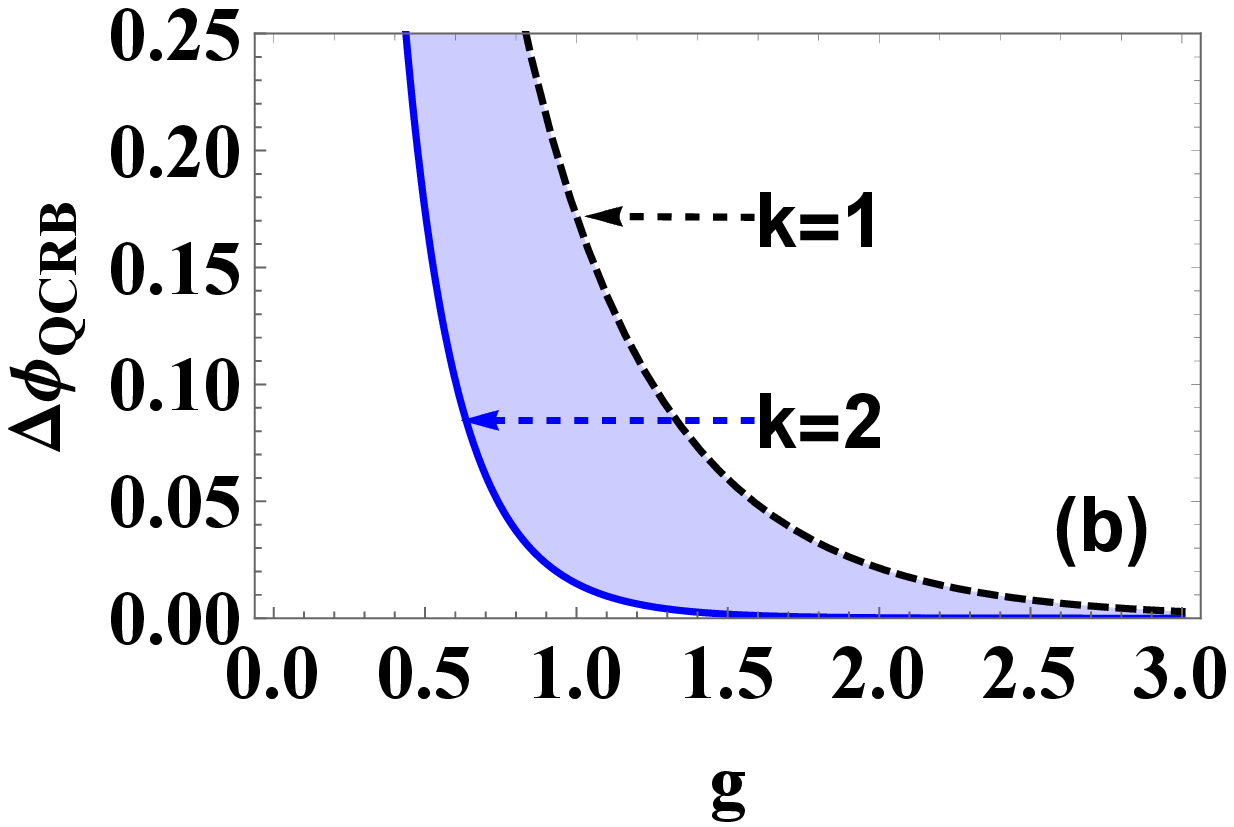} \newline%
\caption{(Color online) (a) The quantum Fisher information $F$ and (b) the
QCRB as a function of gain factor g for $\left \vert \alpha \right \vert =1$ (
$\theta_{\alpha}=\pi/2$), respectively. The black-dashed and blue-solid lines
correspond to linear phase shift ($k=1$) and Kerr nonlinear phase shift
($k=2$), respectively.}%
\end{figure}\begin{figure}[ptb]
\label{Fig7}
\centering \includegraphics[width=0.7\columnwidth]{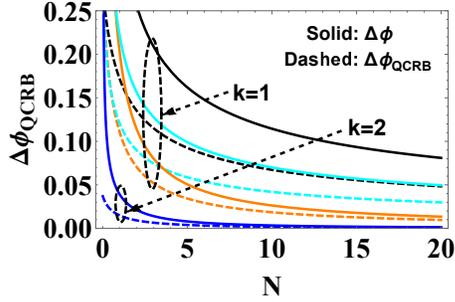}\caption{{}(Color
online) The $\Delta \phi_{QCRB}$ as a function of the total average photon
number N of input state with $g=1.$ The orange and cyan lines correspond to a
coherent state plus a squeezed vacuum state input and two coherent states,
respectively, with the linear phase shift ($k=1$); while the black and blue
lines correspond to $k=1$, and the Kerr nonlinear phase shift ($k=2$),
respectively. The dashed and solid lines correspond to the $\Delta \phi_{QCRB}$
and the $\Delta \phi,$ respectively.}%
\end{figure}

\section{The QFI in the KSU(1,1) interferometer}

As an elegant approach, the QFI can be used to visually describe the maximum
amount of information about the unknown phase shift $\phi$, which is connected
with the QCRB. In fact, the QFI is the intrinsic information in quantum state
and is independent of any specific detection scheme. In the absence of
losses,\ for a pure state, the corresponding QFI can be calculated as
\cite{34,48,49}
\begin{equation}
F=4\left[  \left \langle \psi_{\phi}^{\prime}|\psi_{\phi}^{\prime}\right \rangle
-\left \vert \left \langle \psi_{\phi}^{\prime}|\psi_{\phi}\right \rangle
\right \vert ^{2}\right]  , \label{6}%
\end{equation}
where $\left \vert \psi_{\phi}\right \rangle =\hat{S}\left(  \phi,k\right)
\hat{S}\left(  \xi_{1}\right)  \left \vert \psi_{in}\right \rangle $ is the
state vector prior to the second OPA and $\left \vert \psi_{\phi}^{\prime
}\right \rangle =\partial \left \vert \psi_{\phi}\right \rangle /\partial \phi.$
Then, for the linear phase shift $\left(  k=1\right)  $ and the Kerr nonlinear
phase shift $\left(  k=2\right)  ,$ the QFI can be, respectively, reformed as
\cite{45}%
\begin{equation}
F_{1}=4\left \langle \Delta^{2}\hat{n}\right \rangle ,F_{2}=4\left \langle
\Delta^{2}\hat{n}^{2}\right \rangle , \label{7}%
\end{equation}
where $\hat{n}=\hat{b}^{\dagger}\hat{b}$ is the photon number operator on mode
$b$ and $\left \langle \Delta^{2}\hat{n}^{j}\right \rangle =$ $\left \langle
\bar{\psi}_{out}\right \vert \left(  \hat{n}^{j}\right)  ^{2}\left \vert
\bar{\psi}_{out}\right \rangle -\left \langle \bar{\psi}_{out}\right \vert
\hat{n}^{j}\left \vert \bar{\psi}_{out}\right \rangle ^{2},\left(  j=1,2\right)
$ with the state vector after the first OPA, i.e., $\left \vert \bar{\psi
}_{out}\right \rangle =\hat{S}\left(  \xi_{1}\right)  \left \vert \psi
_{in}\right \rangle .$ Using the normal ordering forms of the operators,
\begin{align}
(\hat{b}^{\dagger}\hat{b})^{4}  &  =\hat{b}^{\dagger4}\hat{b}^{4}+6\hat
{b}^{\dagger3}\hat{b}^{3}+7\hat{b}^{\dagger2}\hat{b}^{2}+\hat{b}^{\dagger}%
\hat{b},\nonumber \\
(\hat{b}^{\dagger}\hat{b})^{2}  &  =\hat{b}^{\dagger2}\hat{b}^{2}+\hat
{b}^{\dagger}\hat{b}, \label{8}%
\end{align}
and the characteristic function method for calculating average, it is ready to
have
\begin{equation}
F_{1}=4\left(  \bar{A}_{2}+\bar{A}_{1}-\bar{A}_{1}^{2}\right)  ,F_{2}=F_{1}+f,
\label{9}%
\end{equation}
with
\begin{align}
f  &  =4\left[  \bar{A}_{4}+6\left(  \bar{A}_{3}+\bar{A}_{2}\right)  -\bar
{A}_{2}\left(  \bar{A}_{2}+2\bar{A}_{1}\right)  \right]  ,\label{9b}\\
\bar{A}_{m}  &  =m!(\sinh^{2m}g)L_{m}(-\left \vert \alpha \right \vert ^{2}%
),m\in \{1,2,3,4\}, \label{10}%
\end{align}
and $L_{m}(\bullet)$ is the Laguerre polynomials. One can refer to Appendix C
for more details about derivations of $\bar{A}_{m}$. Based on Eq. (\ref{9}),
we can give the QCRB $\Delta \phi_{QCRB},$ which represents the lower bound of
the phase sensitivity, i.e.,
\begin{equation}
\Delta \phi_{QCRB}=\frac{1}{\sqrt{\upsilon F_{k}}},\text{ }\left(
k=1,2\right)  , \label{11}%
\end{equation}
where $\upsilon$ is the number of trials. For simplicity, here we set
$\upsilon=1.$ In general, the smaller the values of $\Delta \phi_{QCRB}$, the
higher the phase sensitivity.

To clearly see this point, according to Eqs. (\ref{9}) and (\ref{11}), at a
fixed $\left \vert \alpha \right \vert =1,$ Fig. 6 shows the QFI and the QCRB
changing with $g$ for different $k=1,2.$ It is clear that, for the cases of
$k=1,2,$ the values of the QFI increase significantly with the increase of
$g,$ thereby leading to the more precise phase sensitivity. Moreover, when
given the same gain factor $g$, both the QFI and the QCRB of $k=2$ always
outperform those of $k=1$, which distinctly shows the superiority of the
nonlinear phase shift compared to the linear case. This result originates from
the additional item $f$ in Eq. (\ref{9}), giving rise to the increase of QFI.

As a comparison, we also consider the $\Delta \phi_{QCRB}$ as a function of $N$
for several different input resources in Fig. 7, similar to the phase
sensitivity (in Fig. 5). Some similar results are obtained. For instance,
although the QCRB with $\left \vert \alpha \right \rangle _{a}\otimes \left \vert
0\right \rangle _{b}$ is the worst than that with other resources in the
traditional SU(1,1) interferometer \cite{63}, the smallest $\Delta \phi_{QCRB}$
can be realized by the simplest input $\left \vert \alpha \right \rangle
_{a}\otimes \left \vert 0\right \rangle _{b}$ in the KSU(1,1) interferometer. In
addition, comparing Fig. 5 with Fig. 7, it is found that compared with several
different input resources in the traditional SU(1,1) interferometer, the input
$\left \vert \alpha \right \rangle _{a}\otimes \left \vert 0\right \rangle _{b}$ in
the KSU(1,1) interferometer is closer to the ultimate phase precision
$\Delta \phi_{QCRB}.$

\section{The effects of photon losses}

In the realistic scenario, the decoherence process is unavoidable because
there always exist interaction between the interferometer system and its
surrounding environment, thereby leading to the information leakage from the
system to the environment. For this season, the system performance would drop
severely. In general, there are many types of decoherence process
\cite{64,65}, such as photon loss, imperfect detector, phase diffusion, and so
on. Here, for simplicity, we only study the effects of photon losses on both
the phase sensitivity and the QFI in our scheme.

\begin{figure}[ptb]
\label{Fig8}
\centering \includegraphics[width=1.0\columnwidth]{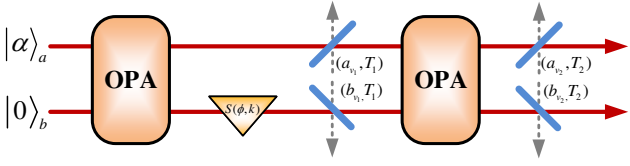}\caption{{}(Color
online) Schematic diagram of the photon losses that occur before and after the
second OPA. $T_{j}\left(  j=1,2\right)  $ is the transmissivity of the
fictitious beam splitters; $a_{v_{j}}$ and $b_{v_{j}}$ are the vacuum noise
operators.}%
\end{figure}

\subsection{The effects of photon losses on the phase sensitivity}

\begin{figure}[ptb]
\label{Fig9}
\centering \includegraphics[width=0.73\columnwidth]{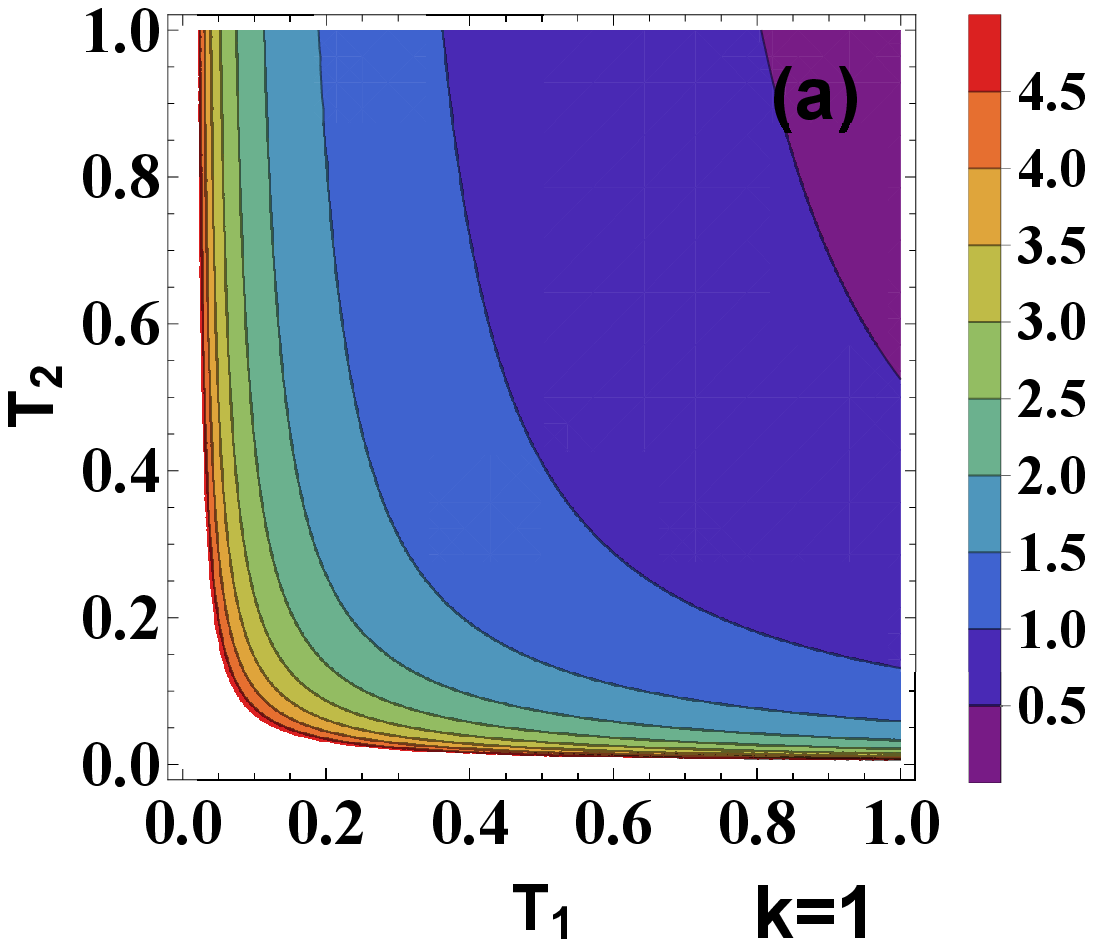}\newline%
\includegraphics[width=0.72\columnwidth]{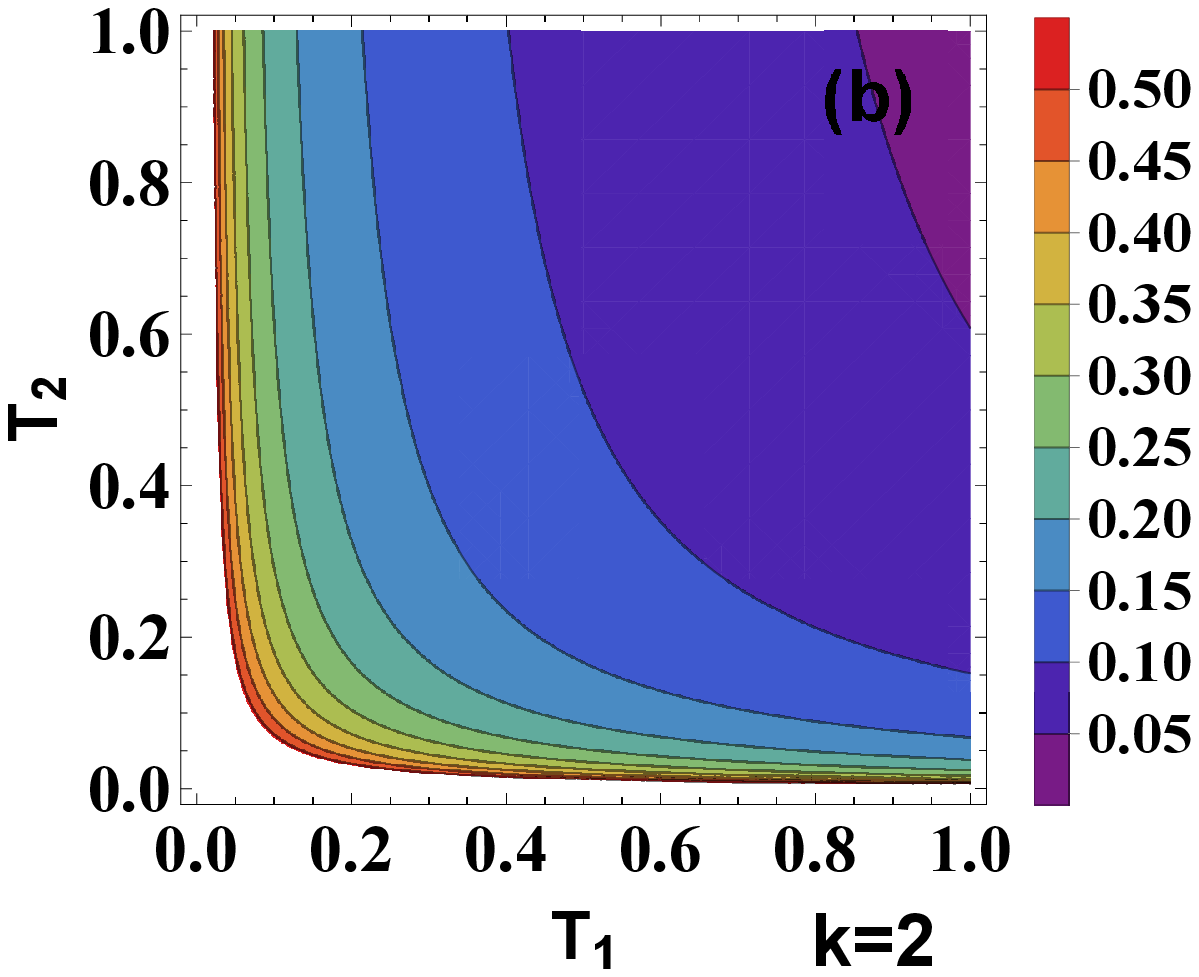} \newline \caption{(Color
online) Phase sensitivity based on homodyne detection as a function of the
photon losses inside the interferometer $T_{1}$ and the photon losses outside
the interferometer $T_{2}$ with $g=1$ and $\left \vert \alpha \right \vert =1$
($\theta_{\alpha}=\pi/2$). (a) and (b) correspond to the cases with$k=1,2$,
respectively.}%
\end{figure}\begin{figure}[ptb]
\label{Fig10}
\centering \includegraphics[width=0.73\columnwidth]{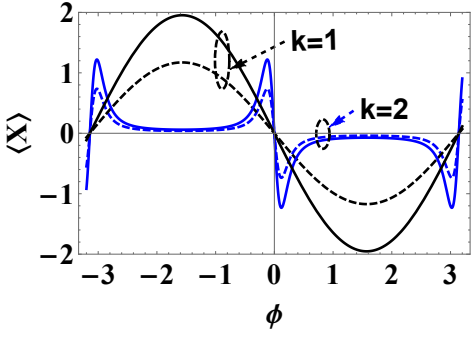}\newline%
\caption{(Color online) The output signal as a function of $\phi$ with $g=1$,
$\left \vert \alpha \right \vert =1$, $\theta_{\alpha}=\frac{\pi}{2},$and
$T_{1}=T_{2}=0.6$. The black and blue lines correspond to $k=1,2$,
respectively. The dashed and solid lines correspond to the effects of photon
losses and no photon losses, respectively.}%
\end{figure}

First, let us consider the effects of photon losses before and after the
second OPA on the phase sensitivity. Actually, the losses after the second OPA
can also be seen as the detection imperfect at out ports. To describe the
photon losses, the channel is usually simulated by inserting the fictitious
beam splitter (BS$_{j}$) with transmissivity $T_{j}\left(  j=1,2\right)  ,$ as
shown in Fig. 8. Generally speaking, the smaller the values of $T_{j}$, the
more serious the photon loss. Here, we denote the photon loss inside and
outside the interferometer as internal loss ($T_{1}$) and external one
($T_{2}$), respectively. In the following discussion, we assume that the
losses of different modes are the same inside or outside the interferometer.

For any single-mode pure state $\left \vert \varphi \right \rangle _{in}$, after
through the photon loss, the output state $\left \vert \varphi \right \rangle
_{out}$ in an enlarged space can be expressed as a pure state, i.e.,
$\left \vert \varphi \right \rangle _{out}=\hat{U}_{BS_{j}}\left \vert
\varphi \right \rangle _{in}\left \vert 0\right \rangle $. Thus, the average value
of operator $\hat{O}$ can be calculated as $\left \langle \hat{O}\right \rangle
=\left.  \left \langle 0\right \vert _{in}\left \langle \varphi \right \vert
\right.  \hat{U}_{BS_{j}}^{\dagger}\hat{O}\hat{U}_{BS_{j}}\left.  \left \vert
\varphi \right \rangle _{in}\left \vert 0\right \rangle \right.  $, which is
equivalent to the average value in the reduced density operator $\rho_{out}%
=$Tr$[\hat{U}_{BS_{j}}\left \vert \varphi \right \rangle _{in}\left \vert
0\right \rangle \left \langle 0\right \vert _{in}\left \langle \varphi \right \vert
\hat{U}_{BS_{j}}^{\dagger}]$. In our scheme in Fig. 8, the final output state
$\left \vert \varphi \right \rangle _{f}$ in the enlarged space is given by%
\begin{equation}
\left \vert \varphi \right \rangle _{f}=\hat{U}_{BS_{2}}^{b}\hat{U}_{BS_{2}}%
^{a}\hat{S}\left(  \xi_{2}\right)  \hat{U}_{BS_{1}}^{b}\hat{U}_{BS_{1}}%
^{a}\left \vert \psi_{\phi}\right \rangle \left \vert \vec{0}\right \rangle ,
\label{12a}%
\end{equation}
where $\left \vert \vec{0}\right \rangle =\left \vert 0000\right \rangle $ is the
noise vacua, and $\hat{U}_{BS_{j}}^{a}=\exp[\arccos \sqrt{T_{j}}(\hat
{a}^{\dagger}\hat{a}_{\upsilon_{j}}-\hat{a}\hat{a}_{\upsilon_{j}}^{\dagger})]$
and $\hat{U}_{BS_{l}}^{b}=\exp[\arccos \sqrt{T_{l}}(\hat{b}^{\dagger}\hat
{b}_{\upsilon_{l}}-\hat{b}\hat{b}_{\upsilon_{l}}^{\dagger})]$ are the BS
operators acting on mode $\hat{a}$ and $\hat{b}$, respectively. Here $\hat
{a}_{\upsilon_{j}}$ and $\hat{b}_{\upsilon_{l}}$ are the vacuum noise operators.

Using the transformations of the fictitious BS$_{j}$ on the annihilation
operators, i.e.,
\begin{align}
(\hat{U}_{BS_{j}}^{a})^{\dagger}\hat{a}\hat{U}_{BS_{j}}^{a} &  =\sqrt{T_{j}%
}\hat{a}+\sqrt{1-T_{j}}\hat{a}_{\upsilon_{j}},\label{12b}\\
(\hat{U}_{BS_{l}}^{b})^{\dagger}\hat{b}\hat{U}_{BS_{l}} &  =\sqrt{T_{l}}%
\hat{b}+\sqrt{1-T_{l}}\hat{b}_{\upsilon_{l}},\label{12}%
\end{align}
it is not difficult to obtain the phase sensitivity $\Delta \phi_{L_{k}}$
($k=1,2$) after losses, not shown here for simplicity. In particular, at the
optimal point $\phi=0,$ for the traditional and KSU(1,1) interferometers in
the presence of photon loss, the phase sensitivities $\Delta \phi_{L_{k}}$
$(k=1,2)$ are derived, respectively, as%
\begin{equation}
\Delta \phi_{L_{1}}=\sqrt{\left(  \Delta \phi_{1}\right)  ^{2}+\frac{\left(
1-T_{1}\right)  T_{2}\cosh2g+1-T_{2}}{4T_{1}T_{2}\left \vert \alpha \right \vert
^{2}\sinh^{4}g}},\label{14}%
\end{equation}
and
\begin{equation}
\Delta \phi_{L_{2}}=\frac{\Delta \phi_{L_{1}}}{1+N_{OPA}\left(  N_{\alpha
}+2\right)  },\label{13}%
\end{equation}
where the phase sensitivity $\Delta \phi_{L_{1}}$ corresponds to that of the
traditional SU(1,1) interferometer in the presence of photon loss. In Eq.
(\ref{14}), the first term $\left(  \Delta \phi_{1}\right)  ^{2}$ can be given
by Eq. (\ref{5}), while the second term is resulted from the internal and
external losses. From Eqs. (\ref{14}) and (\ref{13}), it is obvious that
$\Delta \phi_{L_{2}}>\Delta \phi_{L_{1}}\ $in the presence of losses.
\begin{figure}[ptb]
\label{Fig11}
\centering \includegraphics[width=0.7\columnwidth]{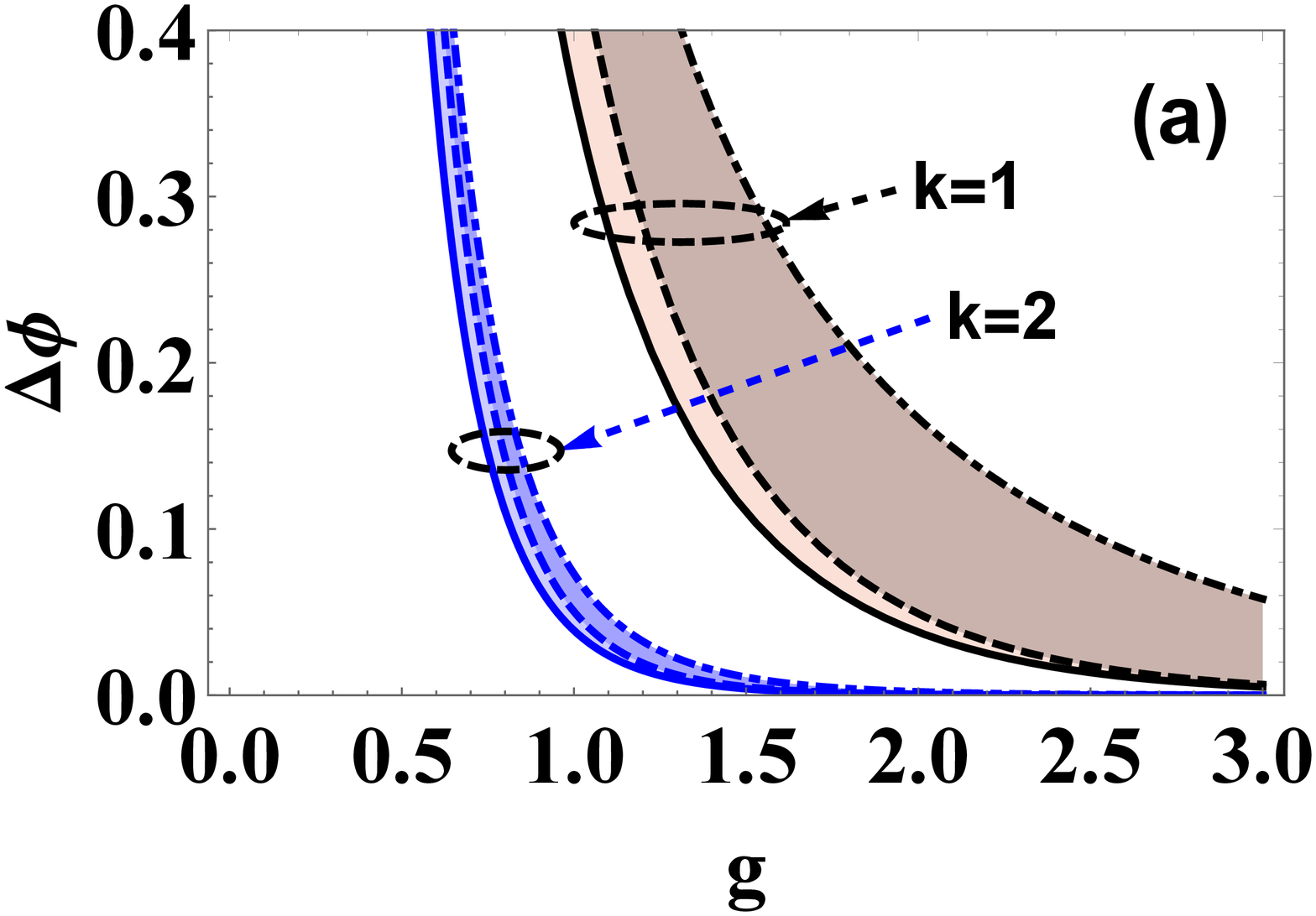}\newline%
\includegraphics[width=0.71\columnwidth]{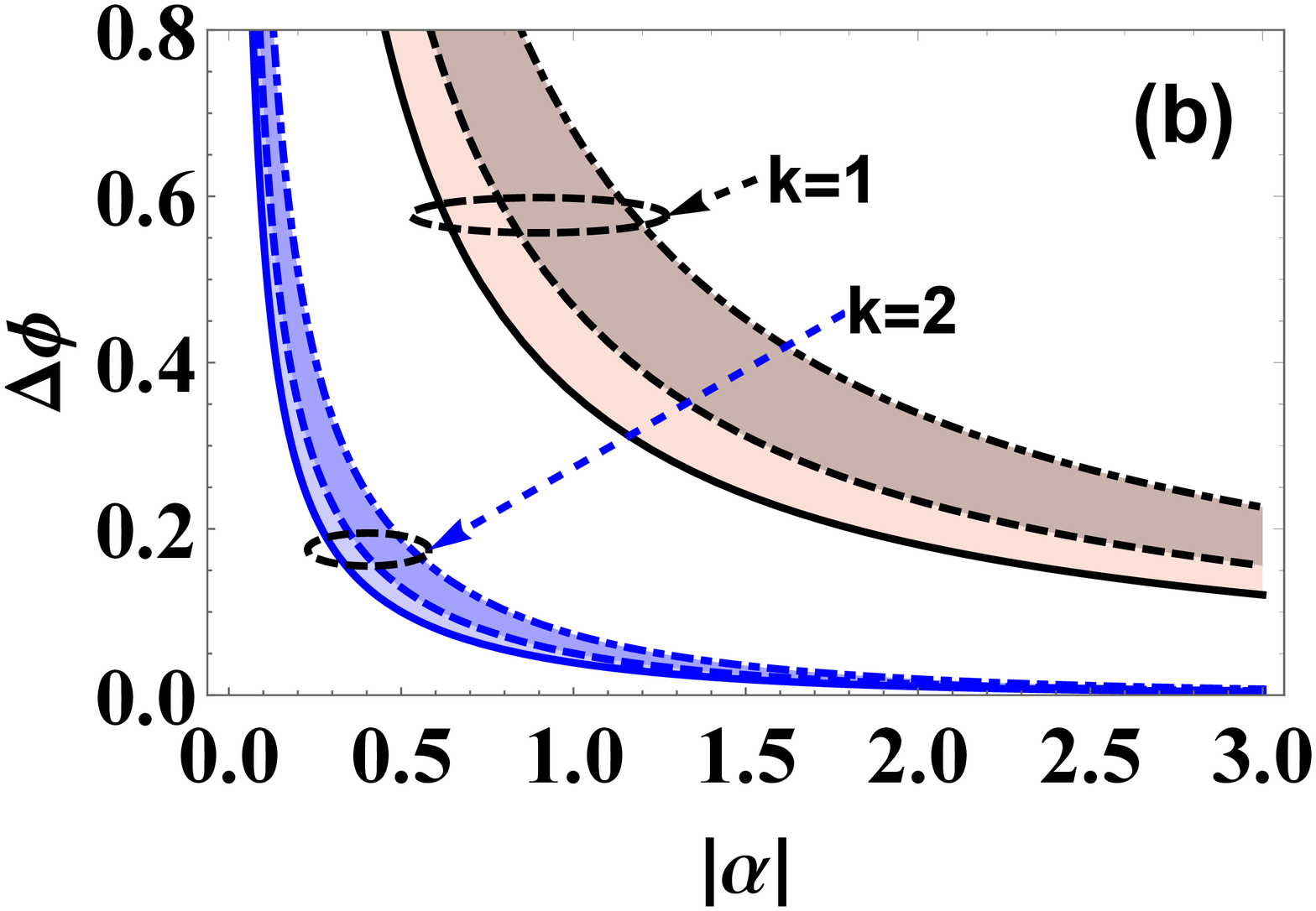} \newline \caption{{}(Color
online) Phase sensitivity based on homodyne detection as a function of (a) the
gain factor $g$ with $\left \vert \alpha \right \vert =1$, (b) the coherent
amplitude $\left \vert \alpha \right \vert $ with $g=1$ $(\theta_{\alpha}%
=\pi/2).$ The black and blue lines correspond to $k=1,2$, respectively. The
dot-dashed line, dashed line, and solid line correspond to the photon losses
inside the interferometer ($T_{1}=0.6$ and $T_{2}=1$), outside the
interferometer ($T_{1}=1$ and $T_{2}=0.6$), and no photon losses ($T_{1}=$
$T_{2}=1$), respectively.}%
\end{figure}

In order to visually see the influence of losses, at fixed values of $g=1$ and
$\left \vert \alpha \right \vert =1,$ we plot the phase sensitivity $\Delta
\phi_{L_{k}}$ $\left(  k=1,2\right)  $ as a function of $T_{j}\left(
j=1,2\right)  $ for different values of $k=1,2,$ as shown in Fig. 9. It is
seen that for $k=1,2,$ the values of $\Delta \phi_{L_{k}}$ increase with the
decrease of $T_{j},$ which shows that the phase sensitivity can be deeply
influenced by the photon losses. Compared to the case of $k=1$ (Fig. 9(a)),
the phase sensitivity of $k=2$ (Fig. 9(b)) is more robust against photon
losses under the same conditions. The reason lies in that the Kerr nonlinear
phase shift $\left(  k=2\right)  $ is beneficial to the increase of the
denominator in the error propagation formula (see Fig. 10), thereby resulting
in the improvement of the phase sensitivity even in the presence of the photon
losses. In addition, from Fig. 10, it is also found that the full width of the
main peaks for $k=2$ would broaden at fixed photon-loss factors $T_{j}=0.6$,
which implies that the signal's super-resolution characteristic is going to be
worse. Even so, the usage of the Kerr nonlinear phase shift performs better
than that of the linear one with respect to increasing the signal's
super-resolution. \begin{figure}[ptb]
\label{Fig12}
\centering \includegraphics[width=0.73\columnwidth]{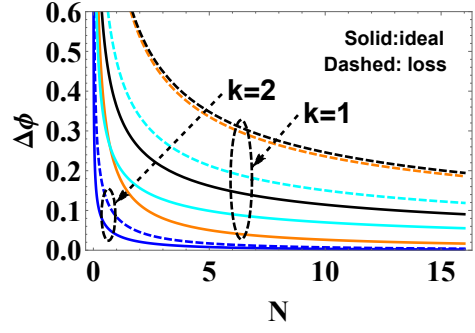}\newline%
\caption{(Color online) Phase sensitivity based on homodyne detection as a
function of the total average photon number N of input state with $g=1,$
$T_{1}=T_{2}=0.6.$ The orange- and cyan-dot-dashed lines respectively
correspond to a squeezed vacuum state plus a coherent states input and two
coherent states with $k=1$, as shown in Ref. \cite{31,32}; The black and blue
lines correspond to $k=1,2$, respectively. The dashed and solid lines
correspond to the effects of photon losses and no photon losses,
respectively.}%
\end{figure}

Next, we consider their own effects of internal and external losses. From Eqs.
(\ref{13}) and (\ref{14}), for the internal and external losses, the phase
sensitivity $\Delta \phi_{L_{1}}$ and $\Delta \phi_{L_{2}}$ can be reduced,
respectively, to
\begin{align}
\left.  \Delta \phi_{L_{1}}\right \vert _{T_{2}=1}  &  =\sqrt{\left(  \Delta
\phi_{1}\right)  ^{2}+\frac{\left(  1-T_{1}\right)  \cosh2g}{4T_{1}\left \vert
\alpha \right \vert ^{2}\sinh^{4}g}},\label{15a}\\
\left.  \Delta \phi_{L_{2}}\right \vert _{T_{2}=1}  &  =\frac{\left.  \Delta
\phi_{L_{1}}\right \vert _{T_{2}=1}}{1+N_{OPA}\left(  N_{\alpha}+2\right)  },
\label{15}%
\end{align}
and
\begin{align}
\left.  \Delta \phi_{L_{1}}\right \vert _{T_{1}=1}  &  =\sqrt{\left(  \Delta
\phi_{1}\right)  ^{2}+\frac{1-T_{2}}{4T_{2}\left \vert \alpha \right \vert
^{2}\sinh^{4}g}},\label{16a}\\
\left.  \Delta \phi_{L_{2}}\right \vert _{T_{1}=1}  &  =\frac{\left.  \Delta
\phi_{L_{1}}\right \vert _{T_{1}=1}}{1+N_{OPA}\left(  N_{\alpha}+2\right)  }.
\label{16}%
\end{align}

From Eqs. (\ref{15a})-(\ref{16}), it is clear that the effects of the internal
losses are greater than those of the external ones on the phase sensitivity
due to $\cosh \left(  2g\right)  >1$ for the same losses. To clearly see this
point, Fig. 11 shows the phase sensitivity $\Delta \phi$ as a function of
$\left \vert \alpha \right \vert $ or $g$ for the internal and external losses.
It is shown that, for a fixed $k$, the gap between the ideal scenario and the
internal-losses one is larger than that between the ideal scenario and the
external-losses one. The gap increases first and then decreases, while the
phase sensitivity still increases as the $\left \vert \alpha \right \vert $ or
$g$ increases. It is interesting that, the gap for $k=2$ becomes significantly
smaller than that for $k=1$, which implies that the Kerr nonlinear phase shift
can effectively resist the influences of photon losses.

Before the end of this part, for some different input states, we make a
comparison about the phase sensitivity in the presence of losses. In Fig. 12,
we plot the $\Delta \phi$ as a function of total average input photon number $N$ for
those resources above, at fixed values $g=1$ and $T_{1}=T_{2}=0.6.$ It is
shown that, for the case of $k=1,$ the gap with $\left \vert \alpha
\right \rangle _{a}\otimes \left \vert \varsigma,0\right \rangle _{b}$ (orange
lines) between the ideal and loss cases is the largest, which means that the
CS plus the SVS as inputs are more sensitive to the photon losses than other
input resources \cite{31}. To some extent, the problem sensitive to noise can
be solved by applying two CSs into the input ports of SU(1,1) interferometer,
see cyan lines in Fig. 12 \cite{32}. The worst performance is still kept by
CS+VS input in lossy channel as in ideal case. However, for the case of $k=2$,
the CS+VS input presents the best performance of phase sensitivity (blue
lines). Even in the presence of photon losses, the phase sensitivity with the
CS+VS is also significantly superior to that with other resources in ideal
case for $k=1$. This indicates that the Kerr nonlinear phase shift can
dramatically suppress the decoherence.

\subsection{The effects of photon losses on the QFI}

For a realistic case, the output state after lossy channel is usually a mixed
state not a pure one. Thus the QFI can not be directly discussed according to
Eq. (\ref{6}). In this case, one may appeal to the spectral decompositions of
density operator \cite{66,67}. Generally speaking, this method is not only
difficult to obtain the spectral decompositions, but also to derive the
analytical expression of the QFI. Fortunately, Escher \textit{et al}.
proposed a way to obtain an upper bound to the QFI in the presence of photon
losses \cite{68}. The basic idea is to purify the whole system involving an
initial pure state and an environment by introducing additional degrees of
freedom, so that the present problem is converted to the parameter estimation
under a unitary evolution $\hat{U}_{S+E}\left(  \phi \right)  $. We shall use
this way to obtain the analytical QFI in a realistic case. For this purpose,
we first make a brief review in the following.

Given an initial pure state $\left \vert \psi_{S}\right \rangle $ in the probe
system $S$ and an initial state $\left \vert 0_{E}\right \rangle $ in the
environment, the purified state $\left \vert \psi_{S+E}\right \rangle $ in the
enlarged space can be given by
\begin{align}
\left \vert \psi_{S+E}\right \rangle  &  =\hat{U}_{S+E}\left(  \phi \right)
\left \vert \psi_{S}\right \rangle \left \vert 0_{E}\right \rangle ,\nonumber \\
&  =\sum_{l=0}^{\infty}\hat{\Pi}_{l}\left(  \phi \right)  \left \vert \psi
_{S}\right \rangle \left \vert l_{E}\right \rangle \label{17}%
\end{align}
where $\hat{\Pi}_{l}\left(  \phi \right)  $ is the Kraus operator describing
the photon losses (also including the effect of phase shift), and $\left \vert
l_{E}\right \rangle $ is an orthogonal basis of the state $\left \vert
0_{E}\right \rangle .$ In this situation, for the whole purified system, the
QFI $C_{Q}\left[  \left \vert \psi_{S}\right \rangle ,\hat{\Pi}_{l}\left(
\phi \right)  \right]  $ turns out to be
\begin{align}
&  C_{Q}\left[  \left \vert \psi_{S}\right \rangle ,\hat{\Pi}_{l}\left(
\phi \right)  \right] \nonumber \\
&  =4\left[  \left \langle \psi_{S+E}^{\prime}|\psi_{S+E}^{\prime}\right \rangle
-\left \vert \left \langle \psi_{S+E}^{\prime}|\psi_{S+E}\right \rangle
\right \vert ^{2}\right]  . \label{18}%
\end{align}

According to Eqs. (\ref{17}) and (\ref{18}), the upper bound $C_{Q}\left[
\left \vert \psi_{S}\right \rangle ,\hat{\Pi}_{l}\left(  \phi \right)  \right]  $
can be given in terms of Kraus operators%
\begin{equation}
C_{Q}\left[  \left \vert \psi_{S}\right \rangle ,\hat{\Pi}_{l}\left(
\phi \right)  \right]  =4\left[  \left \langle \hat{H}_{1}\left(  \phi \right)
\right \rangle _{S}-\left \langle \hat{H}_{2}\left(  \phi \right)  \right \rangle
_{S}^{2}\right]  , \label{20}%
\end{equation}
with the averages $\left \langle \cdot \right \rangle _{S}$ being derived in
$\left \vert \psi_{S}\right \rangle $ and
\begin{align}
\hat{H}_{1}  &  =\sum_{l=0}^{\infty}\frac{d\hat{\Pi}_{l}^{\dagger}\left(
\phi \right)  }{d\phi}\frac{d\hat{\Pi}_{l}\left(  \phi \right)  }{d\phi
},\label{21a}\\
\hat{H}_{2}  &  =i\sum_{l=0}^{\infty}\frac{d\hat{\Pi}_{l}^{\dagger}\left(
\phi \right)  }{d\phi}\hat{\Pi}_{l}\left(  \phi \right)  . \label{21}%
\end{align}
Actually, Eq. (\ref{20}) provides\ an upper bound to the QFI $F_{L}\leqslant
C_{Q}\left[  \left \vert \psi_{S}\right \rangle ,\hat{\Pi}_{l}\left(
\phi \right)  \right]  $ for the reduced system \cite{62}, thus one need to
find the minimal value over all Kraus operators $\left \{  \Pi_{l}\left(
\phi \right)  \right \}  $, i.e.,
\begin{equation}
F_{L}=\underset{\left \{  \Pi_{l}\left(  \phi \right)  \right \}  }{\min}%
C_{Q}\left[  \left \vert \psi_{S}\right \rangle ,\hat{\Pi}_{l}\left(
\phi \right)  \right]  . \label{19}%
\end{equation}
\begin{figure}[ptb]
\label{Fig13}
\centering \includegraphics[width=0.9\columnwidth]{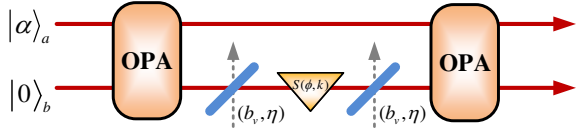}\caption{{}(Color
online) Schematic diagram of the photon losses that occur before and after the
nonlinear phase. $\eta$ is the transmissivity of the fictitious BS; $b_{v}$ is
the vacuum operator.}%
\end{figure}

In our scheme, we consider the QFI of the KSU(1,1) interferometer with the
photon losses placed before or after the Kerr nonlinear phase shift in $b$
arm, as shown in Fig. 13. The corresponding Kraus operator $\hat{\Pi}%
_{l}\left(  \phi \right)  $ including the nonlinear phase can be written as the
general form
\begin{equation}
\hat{\Pi}_{l}\left(  \phi \right)  =\sqrt{\frac{\left(  1-\eta \right)  ^{l}%
}{l!}}e^{i\phi \lbrack(\hat{b}^{\dagger}\hat{b})^{2}-2\mu_{1}\hat{b}^{\dagger
}\hat{b}l-\mu_{2}l^{2}]}\eta^{\frac{\hat{n}}{2}}\hat{b}^{l}, \label{22}%
\end{equation}
where $\eta$ denotes the strength of the photon loss. $\eta=0$ and $\eta=1$
describe the complete absorption and lossless cases, respectively. $\mu_{1}$
and $\mu_{2}$ are two variational parameters with $\mu_{1}=\mu_{2}=0$ and
$\mu_{1}=\mu_{2}=-1$ corresponding to the photon losses occurring before and
after the Kerr nonlinear phase shift, respectively.

To derive Eq. (\ref{20}) using Eq. (\ref{22}), we shall appeal to the
technique of integration within an ordered product of operators (IWOP)
\cite{69} to derive the operator identity [see Appendix D], i.e.,
\begin{equation}
\eta^{\hat{n}}\hat{n}^{q}=\colon \left.  \frac{\partial^{q}}{\partial x^{q}%
}e^{\left(  \eta e^{x}-1\right)  \hat{b}^{\dagger}\hat{b}}\right \vert
_{x=0}\colon, \label{23}%
\end{equation}
where $:\cdot:$ indicates the symbol of the normal ordering form, which
further leads to the formula [see Appendix D]%
\begin{align}
&  \sum_{l=0}^{\infty}\frac{\left(  1-\eta \right)  ^{l}}{l!}l^{p}\hat
{b}^{\dagger l}\eta^{\hat{n}}\hat{n}^{q}\hat{b}^{l}\nonumber \\
&  =D_{q,p}\left[  \eta e^{x}+\left(  1-\eta \right)  e^{y}\right]  ^{\hat
{b}^{\dagger}\hat{b}}, \label{25}%
\end{align}
with $D_{q,p}=\frac{\partial^{q+p}}{\partial x^{q}\partial y^{p}}\left[
\cdot \right]  _{|x=y=0}$ being a partial differential operator.

Using Eq. (\ref{25}) and the following transformation relations
\begin{align}
e^{\lambda \hat{b}^{\dagger}\hat{b}}\hat{b}^{l}e^{-\lambda \hat{b}^{\dagger}%
\hat{b}}  &  =e^{-\lambda l}\hat{b}^{l},\nonumber \\
e^{\lambda(\hat{b}^{\dagger}\hat{b})^{2}}\hat{b}^{l}e^{-\lambda(\hat
{b}^{\dagger}\hat{b})^{2}}  &  =e^{\lambda l^{2}}\hat{b}^{l}e^{-2\lambda
l\hat{b}^{\dagger}\hat{b}}, \label{24}%
\end{align}
the upper bound of the QFI $C_{Q}\left[  \left \vert \psi_{S}\right \rangle
,\hat{\Pi}_{l}\left(  \phi \right)  \right]  $ can be calculated as [see
Appendix E]
\begin{align}
&  C_{Q}\left[  \left \vert \psi_{S}\right \rangle ,\hat{\Pi}_{l}\left(
\phi \right)  \right] \nonumber \\
&  =4(W_{1}^{2}\left \langle \Delta^{2}\hat{n}^{2}\right \rangle -W_{2}%
\left \langle \hat{n}^{3}\right \rangle +W_{3}\left \langle \hat{n}%
^{2}\right \rangle \nonumber \\
&  -W_{4}\left \langle \hat{n}\right \rangle -W_{5}\left \langle \hat{n}%
^{2}\right \rangle \left \langle \hat{n}\right \rangle -W_{6}\left \langle \hat
{n}\right \rangle ^{2}), \label{26}%
\end{align}
where $\left \langle \cdot \right \rangle $ and $\left \langle \Delta^{2}%
\cdot \right \rangle $ are, respectively, the average and variance under the
state $\left \vert \psi_{S}\right \rangle ,$ where $\left \vert \psi
_{S}\right \rangle =S\left(  \xi_{1}\right)  \left \vert \psi_{in}\right \rangle
$ is the state vector after the first OPA. $W_{j}\left(  j=1,2,3,4,5,6\right)
$ are given in Appendix E, not shown here for simplicity. In particular, when
$C_{Q}\left[  \left \vert \psi_{S}\right \rangle ,\hat{\Pi}_{l}\left(
\phi \right)  \right]  $ reaches the minimum value corresponding to the QFI
$F_{L}$, the variational parameters $\mu_{1}$ and $\mu_{2}$ are, respectively,
given by
\begin{align}
\mu_{1opt}  &  =\frac{BE-CD}{AD-2\eta B^{2}},\label{27a}\\
\mu_{2opt}  &  =\frac{AE-2\eta BC}{AD-2\eta B^{2}}, \label{27}%
\end{align}
where $A,B,C,D$ and $E$ are shown in Appendix\ E. Upon substituting those
optimal results $\mu_{1opt}$ and $\mu_{2opt}$ into $C_{Q},$ the minimum value
of $C_{Q}$, i.e., the QFI $F_{L}$ of Kerr nonlinear phase shift in the
presence of photon losses, can be derived theoretically. \begin{figure}[ptb]
\label{Fig14}
\centering \includegraphics[width=0.73\columnwidth]{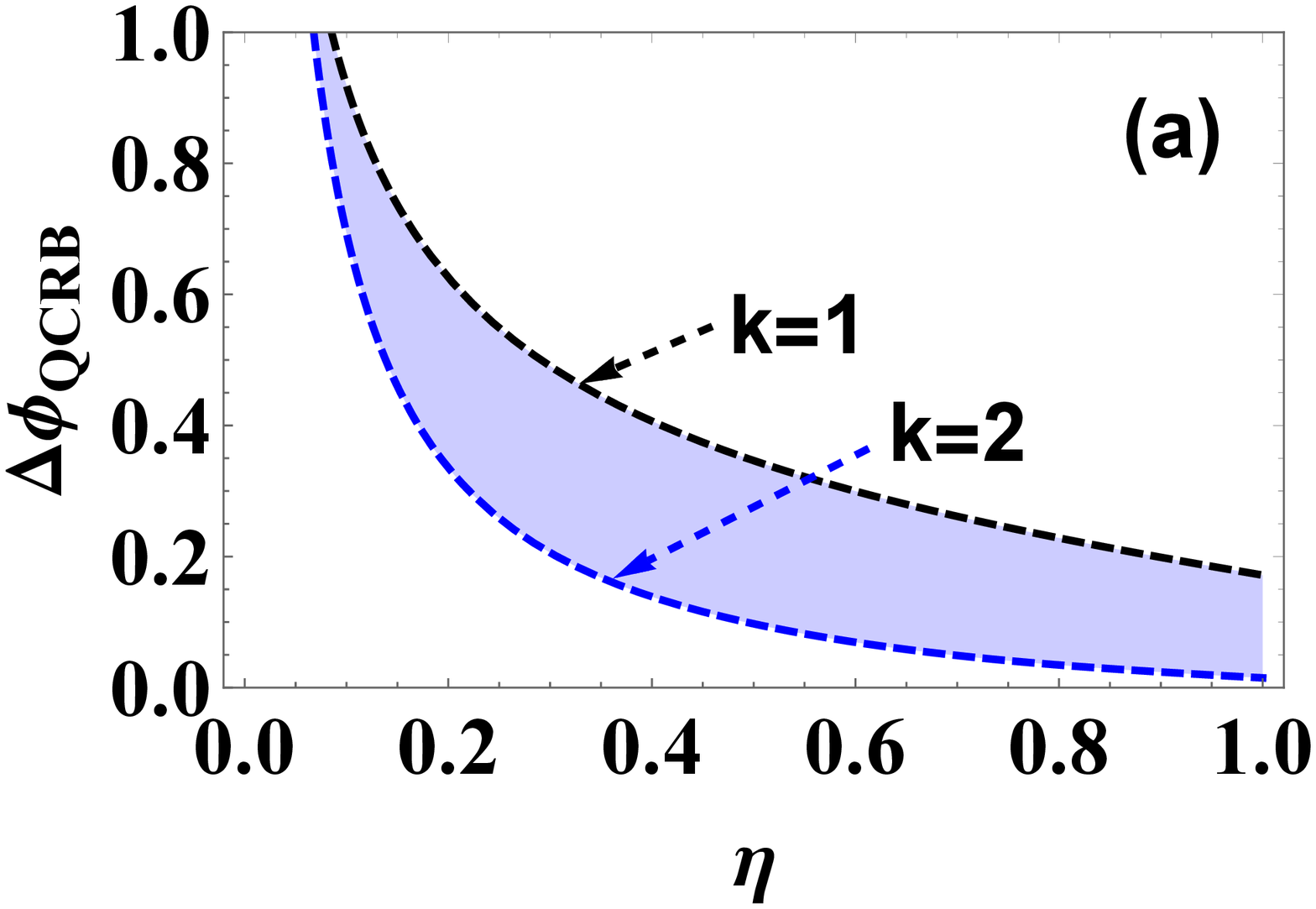}\newline%
\includegraphics[width=0.72\columnwidth]{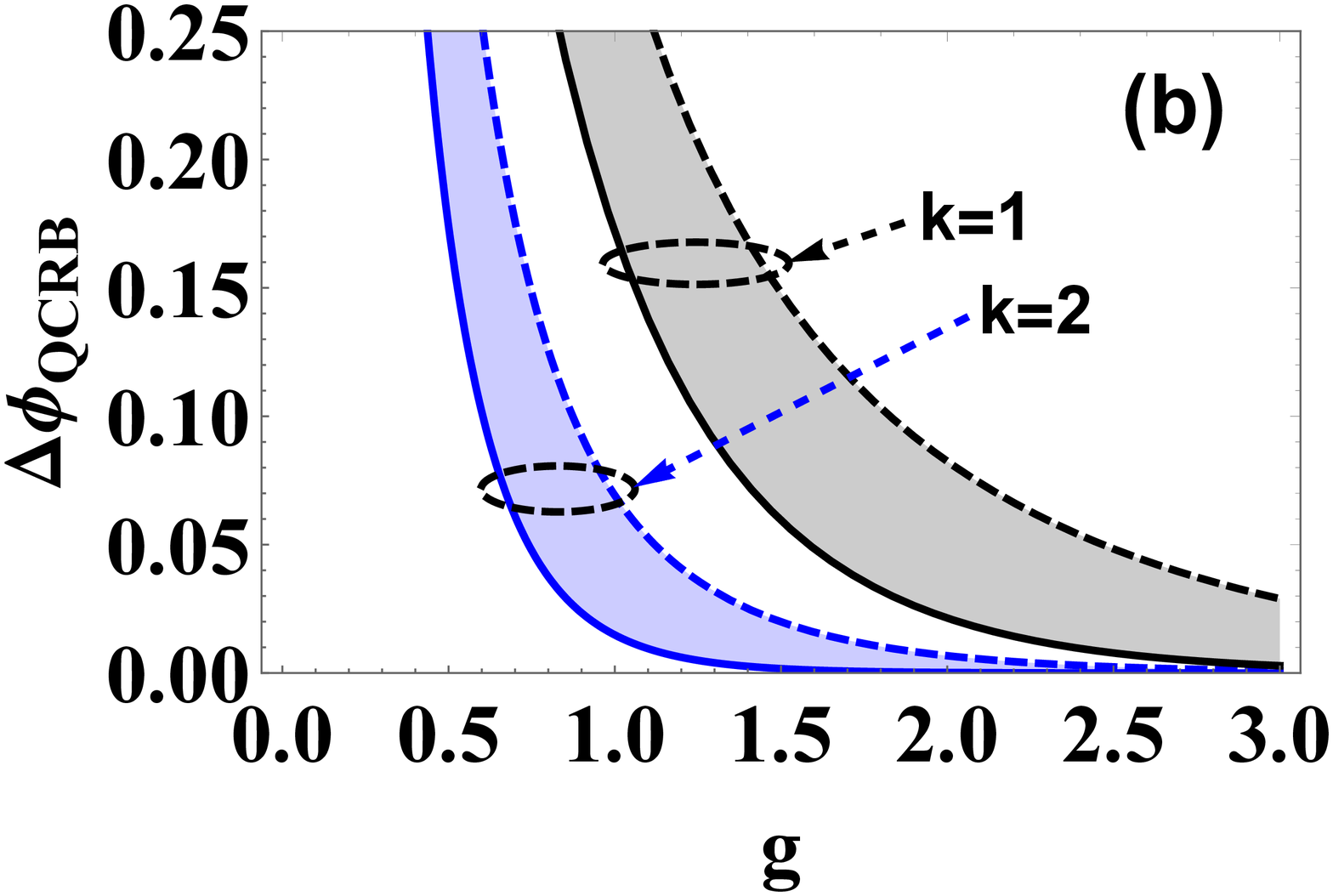} \newline \caption{{}(Color
online) The $\Delta \phi_{QCRB}$ as a function of (a) the transmissivity $\eta$
with $g=1$ and $\left \vert \alpha \right \vert =1$, (b) the gain factor $g$ with
$\left \vert \alpha \right \vert =1.$ The black and blue lines correspond to
$k=1$,$2$, and the dashed and solid lines to the effects of photon losses and
no photon losses, respectively.}%
\end{figure}

Considering a CS and a VS input, here we numerically analyze the QCRB
$\Delta \phi_{QCRB}$ which actually is equivalent to the QFI due to the
relation in Eq. (\ref{11}). For fixed parameters with $|\alpha|=1$ and
$\theta_{\alpha}=\pi/2,$ the QCRB $\Delta \phi_{QCRB}$ as a function of
transmissivity $\eta$ and gain factor $g$ are shown in Fig. 14. For
comparison, the linear case with $k=1$ is also plotted here. It is clear that
the bound performance of phase sensitivity becomes better and better as the
increase of both transmissivity $\eta$ and gain factor $g$ for $k=1,2$ (see
dashed lines in Fig. 14 (a) and (b)). The QCRB for $k=2$ always outperforms
that for $k=1,$ and the gap between them first increase and then decrease as
the increase of $\eta$ and $g$. Compared to the ideal cases (solid lines in
Fig. 14(b)), it is clear that photon losses present an obvious effect on the
QCRB (dashed lines in Fig. 14(b)). However, it is interesting that the gap of
QCRB for $k=2$ between ideal and nonideal cases is significantly smaller than
that for $k=1$, which are also going to be smaller as the increase of $g$
especially for $k=2$. Again, this implies that the combination of Kerr
nonlinear and OPA can further compress the decoherence effect from the
environments.\newline \begin{figure}[ptb]
\label{Fig15}
\centering \includegraphics[width=0.73\columnwidth]{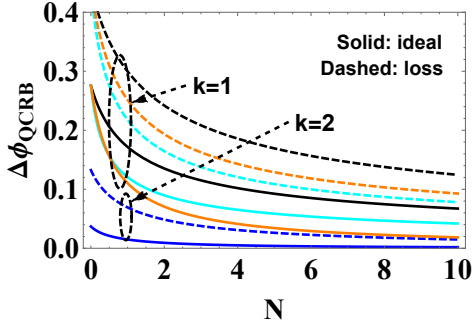}\newline%
\caption{{}(Color online) The $\Delta \phi_{QCRB}$ as a function of the total
average photon number N of input state with $g=1,$ $\eta=0.6.$ The orange- and
cyan-dot-dashed lines respectively correspond to a squeezed vacuum state plus
a coherent states input and two coherent states with $k=1$, as shown in Ref.
\cite{63}. The black and blue lines correspond respectively to $k=1$,2. The
dashed and solid lines correspond to the effects of photon losses and no
photon losses, respectively.}%
\end{figure}

In addition, under the photon-loss processes (e.g., $\eta=0.6$), we further
consider the $\Delta \phi_{QCRB}$ as a function of total average input photon number
$N$ for those different input resources above, when given parameter $g=1,$ as
shown in Fig. 15. Here the solid lines correspond to the ideal cases for a
comparison. It is clearly seen that, compared to other input resources, the
$\left \vert \alpha \right \rangle _{a}\otimes \left \vert 0\right \rangle _{b}$
input presents the worst QCRB in the linear phase shift, with and without
photon losses. However, thanks to the introduction of Kerr nonlinear phase,
the QCRB for $\left \vert \alpha \right \rangle _{a}\otimes \left \vert
0\right \rangle _{b}$ input is the best, which is superior to, even in nonideal
case, the QCRB in ideal case with other inputs. This means that our scheme
with and without photon losses shows an obvious advantage of low-cost input
resources to obtain a better QCRB by introducing nonlinear phase instruments
in the SU(1,1) interferometer system.

\section{Conclusions}

In summary, we propose a protocol of second-order nonlinear phase estimation
by introducing the Kerr medium into the traditional SU(1,1) interferometer. A
kind of simplest inputs, coherent state plus vacuum state, and homodyne
detection for one mode are used. Both the phase sensitivity and the QFI or
QCRB are analytically derived in both ideal and nonideal scenario by using the
CF method and the IWOP technique. It is shown that the increase of both gain
factor $g$ and coherent amplitude $\left \vert \alpha \right \vert $ is
beneficial for improving both the phase sensitivity and the QCRB. Compared to
the linear phase estimation, our scheme presents a significantly better
performance about both of them, especially around the optimal point $\phi$
$=0$. In particular, for the ideal case, our scheme breaks through both the
SQL and the HL, even approaching the SHL, at the large $g$ and $\left \vert
\alpha \right \vert $ levels. In addition, we consider the effects of the
decoherence on the phase sensitivity, from the internal or external photon
losses. It is found that the former has a more obvious decoherence than the
latter. Even so, our scheme still provides a significant improvement for the
performance of phase estimation by homodyne detection, compared to the linear case.

To further show the advantages of our scheme, with and without photon losses,
we also consider both the phase sensitivity and the QCRB changing with total
average input photon number $N$ for several different input resources, including
$\left \vert \alpha \right \rangle _{a}\otimes \left \vert 0\right \rangle _{b},$
$\left \vert \alpha \right \rangle _{a}\otimes \left \vert \beta \right \rangle _{b}$
and $\left \vert \alpha \right \rangle _{a}\otimes \left \vert \varsigma
,0\right \rangle _{b}$, when given parameters $g=1.$ It is found that the best
phase sensitivity can be achieved only using the simplest input $\left \vert
\alpha \right \rangle _{a}\otimes \left \vert 0\right \rangle _{b}$ in the proposed
KSU(1,1) interferometer, which is significantly superior to those inputs in
the traditional SU(1,1) interferometer. This means that, for a simple input
with less energy and less resources, both the phase sensitivity and the QCRB
can be further improved by introducing Kerr nonlinear phase. These results may
have important applications in other quantum information processing.

\begin{acknowledgments}
This work was supported by the National Natural Science Foundation of China
(Grant Nos. 11964013, 11664017), the Training Program for Academic and
Technical Leaders of Major Disciplines in Jiangxi Province (20204BCJL22053),
the Postgraduate Scientific Research Innovation Project of Hunan Province
(Grant No. CX20190126) and the Postgraduate Independent Exploration and
Innovation Project of Central South University (Grant No. 2019zzts070).
\end{acknowledgments}

\textbf{Appendix\ A: Proof of the transformation relation}

For a completeness, here, we give the proof about the transformation relation,
i.e., $\hat{S}^{\dagger}\left(  \phi,2\right)  \hat{b}^{\dagger}\hat{S}\left(
\phi,2\right)  =e^{-i\phi}\hat{b}^{\dagger}e^{-i2\phi \hat{b}^{\dagger}\hat{b}%
}$. It is well known that any operator $\hat{\zeta}$ can be expanded in Fock
state space,
\begin{equation}
\hat{\zeta}=\underset{m,n=0}{\overset{\infty}{\sum}}\zeta_{m,n}\left \vert
m\right \rangle \left \langle n\right \vert , \tag{A1}%
\end{equation}
where $\zeta_{m,n}=\left \langle m\right \vert \hat{\zeta}\left \vert
n\right \rangle $ is the matrix element $\hat{\zeta}$ in Fock space.

Thus, if we take
\begin{align}
\hat{\zeta}  &  =\hat{S}^{\dagger}\left(  \phi,2\right)  \hat{b}^{\dagger}%
\hat{S}\left(  \phi,2\right) \nonumber \\
&  =e^{-i\phi \left(  \hat{b}^{\dagger}\hat{b}\right)  ^{2}}\hat{b}^{\dagger
}e^{i\phi \left(  \hat{b}^{\dagger}\hat{b}\right)  ^{2}}, \tag{A2}%
\end{align}
then the matrix element $\zeta_{m,n}$ can be calculated as
\begin{equation}
\zeta_{m,n}=\sqrt{n+1}e^{-i\phi \left(  2n+1\right)  }\delta_{m,n+1}, \tag{A3}%
\end{equation}
where we have used the $\hat{b}^{\dagger}\hat{b}\left \vert n\right \rangle
=n\left \vert n\right \rangle $ and $\hat{b}^{\dagger}\left \vert n\right \rangle
=\sqrt{n+1}\left \vert n+1\right \rangle .$ Substituting Eq. (A3) into Eq.(A1),
we can get
\begin{align}
\hat{\zeta}  &  =\hat{S}^{\dagger}\left(  \phi,2\right)  \hat{b}^{\dagger}%
\hat{S}\left(  \phi,2\right) \nonumber \\
&  =\underset{m,n=0}{\overset{\infty}{\sum}}\sqrt{n+1}e^{-i\phi \left(
2n+1\right)  }\delta_{m,n+1}\left \vert m\right \rangle \left \langle
n\right \vert \nonumber \\
&  =\underset{n=0}{\overset{\infty}{\sum}}\sqrt{n+1}e^{-i\phi \left(
2n+1\right)  }\left \vert n+1\right \rangle \left \langle n\right \vert
\nonumber \\
&  =e^{-i\phi}\underset{n=0}{\overset{\infty}{\sum}}e^{-i2n\phi}\sqrt
{n+1}\left \vert n+1\right \rangle \left \langle n\right \vert \nonumber \\
&  =e^{-i\phi}\hat{b}^{\dagger}e^{-i2\phi(\hat{b}^{\dagger}\hat{b})}. \tag{A4}%
\end{align}

\textbf{Appendix\ B: Phase estimation based on homodyne measurement}

Combining Eqs. (\ref{2}) and (\ref{3}), we can derive the standard deviation
$\Delta \hat{X}$ as
\begin{equation}
\Delta \hat{X}=\sqrt{\left \vert \overline{U}\right \vert ^{2}+\left \vert
\overline{V}\right \vert ^{2}+\overline{O}}, \tag{B1}%
\end{equation}
where we have set
\begin{equation}
\left \vert \overline{U}\right \vert ^{2}+\left \vert \overline{V}\right \vert
^{2}=\cosh^{2}2g-\text{Re}\left(  e^{i\phi}\overline{I}\right)  \sinh^{2}2g,
\tag{B2}%
\end{equation}
and%
\begin{equation}
\overline{O}=2\left \vert \alpha \right \vert ^{2}\left \vert u\right \vert
^{2}(1-\left \vert \overline{I}\left(  g,\phi \right)  \right \vert
^{2})+2\text{Re}\left(  Z_{1}+Z_{2}\right)  , \tag{B3}%
\end{equation}
with
\begin{align}
u  &  =-e^{-i\phi}\sinh^{2}g,\nonumber \\
\chi \left(  g,\phi \right)   &  =\frac{1}{\cosh^{2}g-e^{i2\phi}\sinh^{2}%
g},\nonumber \\
\overline{I}\left(  g,\phi \right)   &  =\chi^{2}\left(  g,\phi \right)  \exp \{
\left \vert \alpha \right \vert ^{2}\left[  \chi \left(  g,\phi \right)  -1\right]
\},\nonumber \\
Z_{1}  &  =2(\left \vert \alpha \right \vert ^{2}+\alpha^{\ast2})u^{\ast
}\overline{I}\left(  g,\phi \right) \nonumber \\
&  \times \left[  \chi \left(  g,\phi \right)  -1\right]  \cosh^{2}g,\nonumber \\
Z_{2}  &  =\alpha^{2}u^{2}\overline{I}^{\ast}\left(  g,2\phi \right)
\nonumber \\
&  \times \lbrack e^{-i2\phi}\chi^{\ast}\left(  g,2\phi \right)  -\overline
{I}^{\ast}\left(  g,2\phi \right)  ], \tag{B4}%
\end{align}
and the derivative of $\left \langle \hat{X}\right \rangle $
\begin{equation}
\partial \left \langle \hat{X}\right \rangle /\partial \phi=2\text{Re}\left(
Z_{3}Z_{4}\right)  , \tag{B5}%
\end{equation}
where Re denotes the real part, and
\begin{align}
Z_{3}  &  =i\alpha^{\ast}u^{\ast}\overline{I}\left(  g,\phi \right)
,\nonumber \\
Z_{4}  &  =1+4e^{i2\phi}\left \vert u\right \vert \chi \left(  g,\phi \right)
\nonumber \\
&  +2\left \vert \alpha \right \vert ^{2}e^{i2\phi}\left \vert u\right \vert
\chi^{2}\left(  g,\phi \right)  . \tag{B6}%
\end{align}
Substituting Eqs. (B1) and (B5) into the error propagation formula in Eq.
(\ref{4}), the explicit expression of the phase sensitivity can be derived
theoretically. In particular, when $\phi=0,$ we can get $\left \vert
\overline{U}\right \vert ^{2}+\left \vert \overline{V}\right \vert ^{2}=1$ and
$\overline{O}=0.$ Therefore, the standard deviation $\Delta \hat{X}=1.$
Moreover, utilizing the results from Eq. (B5) at the optimal phase point
$\phi=0,$ one can find the absolute value of the derivative%
\begin{align}
&  \text{ \  \ }\left \vert \partial \left \langle \hat{X}\right \rangle
/\partial \phi \right \vert \nonumber \\
&  =\sqrt{N_{\alpha}}N_{OPA}\left[  1+N_{OPA}\left(  N_{\alpha}+2\right)
\right]  \sin \theta_{\alpha}. \tag{B7}%
\end{align}
Then, after taking $\theta_{\alpha}=\frac{\pi}{2}$ leading to $\sin
\theta_{\alpha}=1$, we can obtain Eq. (\ref{5}).

\textbf{Appendix C: The QFI in ideal case}

For pure quantum system, the QFI can be calculated by Eq. (\ref{7}), where the
average value of operator $A_{m}=\hat{b}^{\dagger m}\hat{b}^{m}$ are needed
(see Eq. (\ref{8})). In order to obtain the QFI, here we use the method of
characteristic function (CF). For any two-mode system, the CF is defined as
\begin{equation}
C_{W}\left(  z_{1},z_{2}\right)  =\mathtt{Tr}\left[  \rho_{out}D\left(
z_{1}\right)  D\left(  z_{2}\right)  \right]  , \tag{C1}%
\end{equation}
where $D\left(  z\right)  =\exp \left(  z\hat{a}^{\dagger}-z^{\ast}\hat
{a}\right)  $ is the displacement operator and $\rho_{out}=\left \vert
\varphi_{out}\right \rangle \left \langle \varphi_{out}\right \vert $ is the
density operator after the first OPA. Then the expectation value $\bar{A}%
_{m}=\left \langle \hat{b}^{\dagger m}\hat{b}^{m}\right \rangle $ can be
calculated as
\begin{equation}
\bar{A}_{m}=D_{m}C_{N}\left(  0,z_{2}\right)  , \tag{C2}%
\end{equation}
where $C_{N}\left(  0,z_{2}\right)  =e^{\frac{1}{2}\left \vert z_{2}\right \vert
^{2}}C_{W}\left(  0,z_{2}\right)  $ is the CF corresponding to normal ordering
and $D_{m}=\frac{\partial^{2m}}{\partial z_{2}^{m}\partial \left(  -z_{2}%
^{\ast}\right)  ^{m}}...|_{z_{2}=z_{2}^{\ast}=0}$ is a partial differential
operator. Thus one can use Eq. (C2) to calculate the expectation value
$\bar{A}_{m}.$ For our scheme, the input state $\left \vert \psi_{in}%
\right \rangle =\left \vert \alpha \right \rangle _{a}\otimes \left \vert
0\right \rangle _{b}.$ After going through the first OPA and the phase shift,
the output state is given by $\left \vert \psi_{\phi}\right \rangle =\hat
{S}\left(  \phi,k\right)  \hat{S}\left(  \xi_{1}\right)  \left \vert \psi
_{in}\right \rangle $. Here we should note that these average values of
$\bar{A}_{m}$ are under the state $\hat{S}\left(  \xi_{1}\right)  \left \vert
\psi_{in}\right \rangle $. Then one can obtain
\begin{equation}
\bar{A}_{m}=m!(\sinh^{2m}g)L_{m}(-\left \vert \alpha \right \vert ^{2}), \tag{C3}%
\end{equation}
where we have utilized the relation between Laguerre polynomials and
two-variable Hermit polynomials,
\begin{equation}
L_{m}\left(  xy\right)  =\frac{\left(  -1\right)  ^{m}}{m!}H_{m,m}\left(
x,y\right)  , \tag{C4}%
\end{equation}
and the generating function of $H_{m,m}\left(  x,y\right)  $ is
\begin{align}
&  \text{ \  \ }H_{m,m}\left(  x,y\right) \nonumber \\
&  =\frac{\partial^{2m}}{\partial s^{m}\partial t^{m}}\left.  \exp \left(
-st+sx+ty\right)  \right \vert _{s=t=0}. \tag{C5}%
\end{align}
Thus, substituting Eq. (\ref{8}) and Eq. (C3) into Eq. (\ref{7}), one can get
the explicit expression of the QFI, for the linear phase shift $\left(
k=1\right)  $ and Kerr nonlinear phase shift $\left(  k=2\right)  ,$
respectively,
\begin{align}
F_{1}  &  =4\left(  \bar{A}_{2}+\bar{A}_{1}-\bar{A}_{1}^{2}\right)
,\nonumber \\
F_{2}  &  =F_{1}+f,\nonumber \\
&  f=4\left[  \bar{A}_{4}+6\left(  \bar{A}_{3}+\bar{A}_{2}\right)  -\bar
{A}_{2}\left(  \bar{A}_{2}+2\bar{A}_{1}\right)  \right]  . \tag{C6}%
\end{align}
\textbf{Appendix\ D: The Proof of Eqs. (\ref{23}) and (\ref{25})}

Using the completeness relation of Fock state, one can get%
\begin{align}
&  \text{ \  \ }\eta^{\hat{n}}\hat{n}^{q}\underset{\lambda=0}{\overset{\infty
}{\sum}}\left \vert \lambda \right \rangle \left \langle \lambda \right \vert
\nonumber \\
&  =\sum_{\lambda=0}^{\infty}\eta^{\lambda}\lambda^{q}\frac{\hat{b}%
^{\dagger \lambda}}{\lambda!}\left \vert 0\right \rangle \left \langle
0\right \vert \hat{b}^{\lambda}\nonumber \\
&  =:\underset{\lambda=0}{\overset{\infty}{\sum}}\eta^{\lambda}\lambda
^{q}\frac{1}{\lambda!}\hat{b}^{\dagger \lambda}e^{-\hat{b}^{\dagger}\hat{b}%
}\hat{b}^{\lambda}\colon \nonumber \\
&  =\colon \left.  e^{-\hat{b}^{\dagger}\hat{b}}\underset{\lambda=0}%
{\overset{\infty}{\sum}}\frac{(\eta \hat{b}^{\dagger}\hat{b})^{\lambda}%
}{\lambda!}\frac{\partial^{q}}{\partial x^{q}}e^{x\lambda}\right \vert
_{x=0}\colon \nonumber \\
&  =\colon \left.  \frac{\partial^{q}}{\partial x^{q}}e^{\left(  \eta
e^{x}-1\right)  \hat{b}^{\dagger}\hat{b}}\right \vert _{x=0}\colon, \tag{D1}%
\end{align}
where we have used the normal ordering form of the vacuum projection operator
\begin{equation}
\left \vert 0\right \rangle \left \langle 0\right \vert =\colon e^{-\hat
{b}^{\dagger}\hat{b}}\colon. \tag{D2}%
\end{equation}
Then, using Eq. (D1), one can calculate the following sum operator, i.e.,%
\begin{align}
&  \underset{l=0}{\overset{\infty}{\sum}}\frac{\left(  1-\eta \right)  ^{l}%
}{l!}l^{p}\hat{b}^{\dagger l}\eta^{\hat{n}}\hat{n}^{q}\hat{b}^{l}\nonumber \\
&  =\underset{l=0}{\overset{\infty}{\sum}}\frac{\left(  1-\eta \right)  ^{l}%
}{l!}l^{p}\colon \left.  (\hat{b}^{\dagger}\hat{b})^{l}\frac{\partial^{q}%
}{\partial x^{q}}e^{\left(  \eta e^{x}-1\right)  \hat{b}^{\dagger}\hat{b}%
}\right \vert _{x=0}\colon \nonumber \\
&  =\colon \left.  \underset{l=0}{\overset{\infty}{\sum}}\frac{\left[  \left(
1-\eta \right)  \hat{b}^{\dagger}\hat{b}\right]  ^{l}}{l!}\frac{\partial^{q+p}%
}{\partial x^{q}\partial y^{p}}e^{\left(  \eta e^{x}-1\right)  \hat
{b}^{\dagger}\hat{b}+yl}\right \vert _{x=y=0}\colon \nonumber \\
&  =\colon \left.  \frac{\partial^{q+p}}{\partial x^{q}\partial y^{p}%
}e^{\left[  \eta e^{x}+\left(  1-\eta \right)  e^{y}-1\right]  \hat{b}%
^{\dagger}\hat{b}}\right \vert _{x=y=0}\colon \nonumber \\
&  =\left.  \frac{\partial^{q+p}}{\partial x^{q}\partial y^{p}}[\eta
e^{x}+\left(  1-\eta \right)  e^{y}]^{\hat{b}^{\dagger}\hat{b}}\right \vert
_{x=y=0}. \tag{D3}%
\end{align}
In the last step, we have used the following operator identity about
$e^{\lambda \hat{b}^{\dagger}\hat{b}}$ i.e.,
\begin{equation}
e^{\lambda \hat{b}^{\dagger}\hat{b}}=\colon e^{\left(  e^{\lambda}-1\right)
\hat{b}^{\dagger}\hat{b}}\colon, \tag{D4}%
\end{equation}
to remove the symbol of normal ordering.

\textbf{Appendix E: The} \textbf{specific expression of }$C_{Q}$

Using Eqs. (\ref{20}), (\ref{25}) and Eqs. (\ref{21}) and (\ref{22}), one can
get%
\begin{align}
C_{Q}  &  =4(W_{1}^{2}\left \langle \Delta^{2}\hat{n}^{2}\right \rangle
-W_{2}\left \langle \hat{n}^{3}\right \rangle +W_{3}\left \langle \hat{n}%
^{2}\right \rangle \nonumber \\
&  \text{ \  \  \ }-W_{4}\left \langle \hat{n}\right \rangle -W_{5}\left \langle
\hat{n}^{2}\right \rangle \left \langle \hat{n}\right \rangle -W_{6}\left \langle
\hat{n}\right \rangle ^{2}), \tag{E1}%
\end{align}
where we have set%
\begin{align}
W_{1}  &  =w_{1}\eta^{2}-2w_{2}\eta-\mu_{2},\nonumber \\
W_{2}  &  =2\eta \left(  3w_{1}^{2}\eta^{3}-3w_{3}\eta^{2}-w_{4}\eta
+w_{5}\right)  ,\nonumber \\
W_{3}  &  =\eta \left(  11w_{1}^{2}\eta^{3}-2w_{6}\eta^{2}+w_{7}\eta
-4w_{1}w_{2}\right)  ,\nonumber \\
W_{4}  &  =\eta \left(  6\eta^{3}-12\eta^{2}+7\eta-1\right)  w_{1}%
^{2},\nonumber \\
W_{5}  &  =2\eta \left(  1-\eta \right)  w_{1}W_{1},\nonumber \\
W_{6}  &  =\eta^{2}\left(  1-\eta \right)  ^{2}w_{1}^{2}, \tag{E2}%
\end{align}
as well as%
\begin{align}
w_{1}  &  =1+2\mu_{1}-\mu_{2},\nonumber \\
w_{2}  &  =\mu_{1}-\mu_{2},\nonumber \\
w_{3}  &  =1+2\left(  3\mu_{1}-2\mu_{2}\right) \nonumber \\
&  \text{ \  \  \  \ }+\left(  2\mu_{1}-\mu_{2}\right)  \left(  4\mu_{1}-3\mu
_{2}\right)  ,\nonumber \\
w_{4}  &  =7\mu_{2}-6\mu_{1}+24\mu_{1}\mu_{2}-14\mu_{1}^{2}-9\mu_{2}%
^{2},\nonumber \\
w_{5}  &  =\mu_{2}w_{1}-2w_{2}^{2},\nonumber \\
w_{6}  &  =9+40\mu_{1}-22\mu_{2}+44\mu_{1}^{2}-48\mu_{1}\mu_{2}+13\mu_{2}%
^{2},\nonumber \\
w_{7}  &  =7+40\mu_{1}-26\mu_{2}+52\mu_{1}^{2}-64\mu_{1}\mu_{2}+19\mu_{2}^{2}.
\tag{E3}%
\end{align}
In particular, when $\mu_{1}=\mu_{2}=-1$, one can obtain the expected result,
i.e.,
\begin{equation}
F_{L_{2}}\leqslant C_{Q}=4\left \langle \Delta^{2}\hat{n}^{2}\right \rangle .
\tag{E4}%
\end{equation}
While for $\mu_{1}=\mu_{2}=0,$ corresponding to the photon losses before the
Kerr nonlinear phase shift, one can get the upper bound, i.e.,
\begin{align}
F_{L_{2}}  &  \leqslant C_{Q}=4[\eta^{4}\left \langle \Delta^{2}\hat{n}%
^{2}\right \rangle +6\eta^{3}\left(  1-\eta \right)  \left \langle \hat{n}%
^{3}\right \rangle \nonumber \\
&  \text{ \  \  \ }+\eta^{2}\left(  11\eta^{2}-18\eta+7\right)  \left \langle
\hat{n}^{2}\right \rangle \nonumber \\
&  \text{ \  \  \ }-\eta \left(  6\eta^{3}-12\eta^{2}+7\eta-1\right)
\left \langle \hat{n}\right \rangle \nonumber \\
&  \text{ \  \  \ }-2\eta^{3}\left(  1-\eta \right)  \left \langle \hat{n}%
^{2}\right \rangle \left \langle \hat{n}\right \rangle \nonumber \\
&  \text{ \  \  \ }-\eta^{2}\left(  1-\eta \right)  ^{2}\left \langle \hat
{n}\right \rangle ^{2}], \tag{E5}%
\end{align}
as expected.

In order to minimize $C_{Q},$ one can take $\partial C_{Q}/\partial \mu
_{1}=\partial C_{Q}/\partial \mu_{2}=0$ for this purpose. Using Eqs. (E1)-(E3),
it is not difficult to obtain two optimization parameters $\mu_{1opt}$ and
$\mu_{2opt}$, which are, respectively, given by
\begin{equation}
\mu_{1opt}=\frac{BE-CD}{AD-2\eta B^{2}},\mu_{2opt}=\frac{AE-2\eta BC}{AD-2\eta
B^{2}}, \tag{E6}%
\end{equation}
where we have set $A=2B_{1}H,$ $B=B_{2}H,$ $C=B_{3}H,$ $D=B_{4}H,$ and $E=\eta
B_{5}H.$ Here the column matrix $H$ and row matrix $B_{j}\left(
j=1,2,3,4,5\right)  $ are defined as the following
\begin{align}
H  &  =(\left \langle \Delta^{2}\hat{n}^{2}\right \rangle ,\left \langle \hat
{n}^{3}\right \rangle ,\left \langle \hat{n}^{2}\right \rangle ,\nonumber \\
&  \text{ \  \  \ }\left \langle \hat{n}\right \rangle ,\left \langle \hat{n}%
^{2}\right \rangle \left \langle \hat{n}\right \rangle ,\left \langle \hat
{n}\right \rangle ^{2})^{T},\nonumber \\
A_{1}  &  =\eta-1,\nonumber \\
A_{2}  &  =6\eta^{2}-6\eta+1,\nonumber \\
A_{3}  &  =11\eta^{2}-11\eta+2,\nonumber \\
A_{4}  &  =2\eta-1,\nonumber \\
B_{1}  &  =\left(  \eta A_{1},-A_{2},A_{3},-A_{2},2\eta A_{1},-\eta
A_{1}\right)  ,\nonumber \\
B_{2}  &  =(A_{1}^{2},-3A_{1}A_{4},A_{3}-A_{4},\nonumber \\
&  \text{ \  \  \ }-A_{2},A_{1}A_{4},-\eta A_{1}),\nonumber \\
B_{3}  &  =\left(  \eta^{2},-3\eta A_{4},A_{3}+A_{4},-A_{2},\eta A_{4},-\eta
A_{1}\right)  ,\nonumber \\
B_{4}  &  =\eta(\eta^{-1}A_{1}^{3},-6A_{1}^{2},A_{3}-2A_{4},\nonumber \\
&  \text{ \  \  \ }-A_{2},2A_{1}^{2},-\eta A_{1}),\nonumber \\
B_{5}  &  =\left(  \eta A_{1},-A_{2},A_{3},-A_{2},\eta^{2}+A_{1}^{2},-\eta
A_{1}\right)  , \tag{E7}%
\end{align}
where the average value $\left \langle \cdot \right \rangle $ is in the state
after the first OPA. Here we should note that, only $H$ are dependent on the
input state, and the other quantities (such as $A_{j},B_{j}$) are independent
of the input state. If $C_{Q}$ can take the minimum value, then it is also the
QFI $F_{L_{2}}$ of Kerr nonlinear phase shift in the presence of photon
losses. In our scheme, if we choose the states $\left \vert \psi_{in}%
\right \rangle =\left \vert \alpha \right \rangle _{a}\otimes \left \vert
0\right \rangle _{b}$ as the inputs of the KSU(1,1) interferometer, then the
states after the first OPA is $\hat{S}\left(  \xi_{1}\right)  \left \vert
\psi_{in}\right \rangle $. Thus the column matrix $H$ can be calculated as
\begin{align}
H  &  =[\frac{F_{2}}{4},\left(  \bar{A}_{3}+3\bar{A}_{2}+\bar{A}_{1}\right)
,\nonumber \\
&  \text{ \  \  \ }\left(  \bar{A}_{2}+\bar{A}_{1}\right)  ,\bar{A}_{1},\bar
{A}_{1}\left(  \bar{A}_{2}+\bar{A}_{1}\right)  ,\bar{A}_{1}^{2}]^{T}, \tag{E8}%
\end{align}
where $F_{2}$ is lossless result in Eq. (\ref{9}) and $\bar{A}_{j}\left(
j=1,2,3\right)  $ and $W_{l}(l=1\sim6)$ can be obtained from Eqs. (C3) and
(E2), respectively.

Further substituting Eq. (E8) and Eq. (E6) into Eq. (E1), the upper bound to
the QFI $C_{Q}\left(  \left \vert \alpha \right \rangle _{a}\otimes \left \vert
0\right \rangle _{b}\right)  $ in our scheme can be obtained as%
\begin{align}
&  \text{ \  \  \ }C_{Q}\left(  \left \vert \alpha \right \rangle _{a}%
\otimes \left \vert 0\right \rangle _{b}\right) \nonumber \\
&  =W_{1}^{2}F_{2}-4[W_{2}\left(  \bar{A}_{3}+3\bar{A}_{2}+\bar{A}_{1}\right)
\nonumber \\
&  \text{ \  \  \ }-W_{3}\left(  \bar{A}_{2}+\bar{A}_{1}\right)  +W_{4}\bar
{A}_{1}\nonumber \\
&  \text{ \  \  \ }+W_{5}\bar{A}_{1}\left(  \bar{A}_{2}+\bar{A}_{1}\right)
+W_{6}\bar{A}_{1}^{2}], \tag{E9}%
\end{align}
which is just the analytical expression of the QFI. In $W_{l}(l=1\sim6)$, the
variational parameters $\mu_{1}$ and $\mu_{2}$ should be replaced,
respectively, by%
\begin{align}
\mu_{1opt}\left(  \left \vert \alpha \right \rangle _{a}\otimes \left \vert
0\right \rangle _{b}\right)   &  =\frac{BE-CD}{AD-2\eta B^{2}},\nonumber \\
\mu_{2opt}\left(  \left \vert \alpha \right \rangle _{a}\otimes \left \vert
0\right \rangle _{b}\right)   &  =\frac{AE-2\eta BC}{AD-2\eta B^{2}}, \tag{E10}%
\end{align}
where $A=2B_{1}H,$ $B=B_{2}H,$ $C=B_{3}H,$ $D=B_{4}H,$ and $E=\eta B_{5}H$
with the column matrix $H$ is shown in Eq. (E8).

In addition, for a comparison between the linear phase shift and the nonlinear
one (our scheme), here we give the QFI with the linear phase shift in the
presence of photon losses, for several different input states, including
$\left \vert \alpha \right \rangle _{a}\otimes \left \vert 0\right \rangle _{b}$
(CS\&VS)$,$ $\left \vert \alpha \right \rangle _{a}\otimes \left \vert
\beta \right \rangle _{b}$ (CS\&CS) and $\left \vert \alpha \right \rangle
_{a}\otimes \left \vert \varsigma,0\right \rangle _{b}$ (CS\&SVS). In a similar
way to derive Eq. (E9), for these states above together with the linear phase
shift, the QFI $F_{L_{1}}$ can be calculated as \cite{68}%

\begin{align}
&  \text{ \  \ }F_{L_{1}\left(  CS\&VS\right)  }\nonumber \\
&  =\frac{4\eta F_{1}\bar{A}_{1}}{\left(  1-\eta \right)  F_{1}+4\eta \bar
{A}_{1}},\nonumber \\
&  \text{ \  \ }F_{L_{1}\left(  CS\&CS\right)  }\nonumber \\
&  =\frac{4\eta F_{\left(  CS\&CS\right)  }\left \langle \hat{n}\right \rangle
_{\left(  CS\&CS\right)  }}{\left(  1-\eta \right)  F_{\left(  CS\&CS\right)
}+4\eta \left \langle \hat{n}\right \rangle _{\left(  CS\&CS\right)  }%
},\nonumber \\
&  \text{ \  \ }F_{L_{1}\left(  CS\&SVS\right)  }\nonumber \\
&  =\frac{4\eta F_{\left(  CS\&SVS\right)  }\left \langle \hat{n}\right \rangle
_{\left(  CS\&SVS\right)  }}{\left(  1-\eta \right)  F_{\left(  CS\&SVS\right)
}+4\eta \left \langle \hat{n}\right \rangle _{\left(  CS\&SVS\right)  }},
\tag{E11}%
\end{align}
where $\left \langle \hat{n}\right \rangle _{\left(  CS\&CS\right)  }$ and
$\left \langle \hat{n}\right \rangle _{\left(  CS\&SVS\right)  }$ are given by
\begin{align}
&  \text{ \  \ }\left \langle \hat{n}\right \rangle _{\left(  CS\&CS\right)
}\nonumber \\
&  =\left(  \left \vert \alpha \right \vert \sinh g+\left \vert \beta \right \vert
\cosh g\right)  ^{2}+\sinh^{2}g,\nonumber \\
&  \text{ \  \ }\left \langle \hat{n}\right \rangle _{\left(  CS\&SVS\right)
}\nonumber \\
&  =\left(  \left \vert \alpha \right \vert ^{2}+1\right)  \sinh^{2}g+\cosh
^{2}g\left(  \sinh^{2}r\right)  , \tag{E12}%
\end{align}
and $F_{\left(  CS\&CS\right)  }$ and $F_{\left(  CS\&SVS\right)  }$ are the
lossless results which can be seen from Ref. \cite{63}, i.e.,%
\begin{align}
&  \text{ \  \ }F_{\left(  CS\&CS\right)  }\nonumber \\
&  =(\left \vert \alpha \right \vert ^{2}+\left \vert \beta \right \vert ^{2}%
)\cosh4g+\sinh^{2}(2g)\nonumber \\
&  \text{ \  \ }+2\left \vert \alpha \right \vert \left \vert \beta \right \vert
\sinh4g+\left \vert \alpha \right \vert ^{2}+\left \vert \beta \right \vert
^{2}\nonumber \\
&  \text{ \  \ }-2\left(  \left \vert \alpha \right \vert ^{2}-\left \vert
\beta \right \vert ^{2}\right)  \cosh2g,\tag{E13}\\
&  \text{ \  \ }F_{\left(  CS\&SVS\right)  }\nonumber \\
&  =\cosh^{2}\left(  2g\right)  [\frac{1}{2}\sinh^{2}\left(  2r\right)
+\left \vert \alpha \right \vert ^{2}]\nonumber \\
&  \text{ \  \ }+\sinh^{2}(2g)(\left \vert \alpha \right \vert ^{2}e^{2r}%
+\cosh^{2}r)\nonumber \\
&  \text{ \  \ }+\left \vert \alpha \right \vert ^{2}\left(  1-2\cosh2g\right)
\nonumber \\
&  \text{ \  \ }+\frac{1}{4}(\cosh4r-1)(2\cosh2g+1). \tag{E14}%
\end{align}

\end{document}